\documentclass[11pt]{article}
\usepackage{jheppub} % HJEP/JHEP style

\makeatletter
\renewcommand{\@fpheader}{}   % remove "Prepared for submission to JHEP"
\makeatother

\usepackage[utf8]{inputenc}
\usepackage[T1]{fontenc}
\usepackage{lmodern}
\usepackage{amsmath,amssymb,amsfonts,amsthm,mathtools}
\usepackage{bm}
\usepackage{microtype}
\usepackage{enumitem}
\usepackage{tikz-cd}
\usepackage{mathrsfs}
\usepackage{thmtools}
\usepackage{hyperref}
\newcommand{\calO}{\mathcal{O}}
\newcommand{\calA}{\mathcal{A}}

\newcommand{\Hom}{\mathrm{Hom}}
\newcommand{\Mat}{\mathrm{Mat}}
\newcommand{\Aut}{\mathrm{Aut}}

\newcommand{\sgn}{\operatorname{sgn}}
\newcommand{\Res}{\mathrm{Res}}

\newcommand{\Gm}{\mathbb G_m}

\newcommand{\PGL}{\mathrm{PGL}}

\newcommand{\Gr}{\mathrm{Gr}}

\newcommand{\Flags}{\mathrm{Flags}}

\newcommand{\X}{\mathcal{X}}
\newcommand{\A}{\mathcal{A}}
\newcommand{\Z}{\mathbb{Z}}
\newcommand{\D}{\mathcal{D}}

\newcommand{\Rbb}{\mathbb{R}}

\newcommand{\body}{\mathfrak{B}}
\newcommand{\Xscr}{\mathscr{X}}
\newcommand{\Ascr}{\mathscr{A}}

\newcommand{\De}{\Delta}
\newcommand{\Br}{\mathcal B_r}

\newcommand{\Ad}{\mathrm{Ad}}
\newcommand{\End}{\mathrm{End}}

\newcommand{\Ch}{\mathrm{Ch}}
\newcommand{\Irr}{\mathrm{Irr}}

\newcommand{\Li}{\mathrm{Li}}

\title{Super Higher--Teichmüller Geometry and Loop Amplitudes}
\author[a]{Chaoming Song}
\affiliation[a]{Department of Physics, University of Miami}
\emailAdd{c.song@miami.edu}

\begin{document}

\abstract{
We construct a supersymmetric extension of the Fock–Goncharov cluster ensemble associated with a split basic classical Lie supergroup \(G\) and a marked bordered surface \(S\). The resulting structure defines a super higher–Teichmüller geometry: a split super-thickening of \((\Ascr_{G,S},\Xscr_{G,S})\) equipped with a mutation atlas preserving a canonical super log-symplectic form. Each super seed carries an integer weight matrix \(W\) encoding Cartan weights of an abelian odd slice, transforming by the column \(g\)-vector rule and giving rise to a flat logarithmic superconnection and a canonical super volume form. On this geometric foundation we define a canonical logarithmic superform \(\Omega_{\mathrm{super}}^{(L)}\) on a loop fibration \(\pi_L:\Xscr^{(L)}_{G,S}\!\to\!\Xscr_{G,S}\) as the relative lift of the base super volume. For $G=PGL(4|4)$, the corresponding super period \(\mathcal P_{\mathrm{super}}=\int_{\mathcal C}\Omega_{\mathrm{super}}^{(L)}\) encodes the loop amplitude data of planar \(\mathcal N=4\) super Yang–Mills, expressed through a unified and triangulation-independent formula that satisfies Steinmann and cluster adjacency, with the even sector given by Chen iterated integrals and the odd sector captured by an invariant BCFW delta.
}

\maketitle

%\tableofcontents

% =========================
\section{Introduction}
% =========================
The geometry underlying scattering amplitudes exhibits deep connections between quantum field theory and the theory of cluster and Teichmüller moduli. The classical decorated and higher–Teichmüller theories of Penner~\cite{Penner1987Decorated} and Fock–Goncharov~\cite{FockGoncharov2006} describe moduli of local systems and provide a natural language for positive and cluster varieties~\cite{Scott2006GrassmannianCluster,FockGoncharov2007Quantization}. On the physics side, the amplituhedron program~\cite{ArkaniHamed2014Amplituhedron,ArkaniHamed2017PositiveGeomCF,ArkaniHamed2012PosGrass} and the positive–geometry formulation of planar $\mathcal N{=}4$ super Yang–Mills amplitudes~\cite{Drummond2008WilsonLoopDuality,Hodges2009MomentumTwistors,Goncharov:2010jf,Dixon:2011nj,Bourjaily:2017bsb,CaronHuot2019HexagonFunctionBootstrap,Spiering:2024jok} have revealed a geometric origin for analytic structures through canonical differential forms. Recent progress for the $m=4$ amplituhedron makes this correspondence explicit: Even-Zohar and collaborators~\cite{even2023cluster,even2024cluster} showed that BCFW recursions generate cluster-adjacent tilings of $\Gr_{4,n}$ and that each tile coincides with the positive part of a cluster variety, whose canonical form is expressed directly in cluster coordinates. Yet a fully supersymmetric and loop-level geometric framework incorporating these structures has remained elusive. The aim of this work is to construct such a framework by extending higher–Teichmüller geometry to the supersymmetric setting and identifying its canonical object, the super period, with loop-level scattering data.

The construction begins with a super higher–Teichmüller moduli space extending the cluster–Poisson moduli $\Xscr_{G_{\bar0},S}$ of a marked bordered surface $S$ and a split Lie group $G_{\bar0}$ to a super $\X$-variety $\Xscr_{G,S}$ for any split basic classical Lie supergroup $G$ with even body $G_{\bar0}$. Each super seed augments the classical exchange data by an integer weight matrix $W$ encoding the Cartan weights of an abelian odd slice of $\mathfrak g_{\bar1}$. The matrix $W$ transforms under mutations by the column $g$-vector rule, defining a horizontal odd frame in which the canonical super log-symplectic two-form and a flat logarithmic superconnection become manifest. The resulting structure defines a mutation-invariant log-canonical super-Poisson geometry whose even body reproduces the classical cluster ensemble, while the odd sector is globally organized by a mutation-covariant lattice of weight gradings represented seedwise by $W$, unique up to supergauge equivalence.

This framework constitutes a genuine super cluster ensemble, a direct generalization of the Fock–Goncharov construction~\cite{FockGoncharov2006,FockGoncharov2007Quantization} to the supersymmetric setting, providing the moduli-theoretic and symplectic foundation for the super symplectic double, a flat logarithmic superconnection, and a canonical logarithmic super volume form. Earlier super–Teichmüller models follow a distinct path: the decorated and quantized formulations of Penner–Zeitlin and collaborators~\cite{penner2019decorated,AghaeiPawelkiewiczTeschner2015} fix a Weil–Petersson-type invariant two-form and describe super–Fuchsian representations of low-rank orthosymplectic groups, without extending the full cluster ensemble or its canonical form. Similarly, algebraic super-cluster approaches~\cite{ovsienko2019cluster,shemyakova2023super,li2021introduction} focus on super Plücker and super Ptolemy relations in the coordinate ring, while the present approach is moduli-theoretic and symplectic, providing a seedwise flat logarithmic superconnection, a globally defined canonical super two-form, and mutation rules preserving both. Together these structures realize a unified super higher–Teichmüller geometry naturally suited to supersymmetric field theory.

On this geometric foundation, a loop fibration $\pi_L:\Xscr^{(L)}_{G,S}\!\to\!\Xscr_{G,S}$ is introduced to encode the $L$-loop directions, together with a boundary quotient describing physical external legs parametrized by boundary minors $\Delta_b(X)$. On this space there exists a canonical logarithmic superform $\Omega_{\mathrm{super}}^{(L)}$, understood as the relative lift of the base canonical super volume form on $\Xscr_{G,S}$; it is covariantly flat under the loop superconnection and logarithmic along a super divisor $D^{(L)}_{\mathrm{super}}$. Its residues enforce Steinmann factorization and operadic gluing on surfaces, and its cohomology class is mutation-invariant, ensuring triangulation independence. The physically relevant observable is the canonical super period
\[
\mathcal P_{\mathrm{super}} = \int_{\mathcal C} \Omega_{\mathrm{super}}^{(L)},
\]
defined over a positive integration cycle $\mathcal C$. This single, triangulation-independent object unifies the analytic and supersymmetric data of loop amplitudes: the even part governs the singularity structure, while the odd delta factor organizes the supersymmetric content of the integrand. For planar $\mathcal N{=}4$ super Yang–Mills theory on disks $D_n$, identifying the boundary minors $\Delta_b(X)$ with momentum–twistor Plücker coordinates~\cite{Hodges2009MomentumTwistors} reduces multi-loop scattering data to the evaluation of this super period. The odd sector reproduces the invariant BCFW delta~\cite{britto2005direct,britto2005new}, while the even sector is realized as a Chen iterated integral, a refined vertical period on a positive fiber curve encoding the loop dependence. The resulting expression for the IR-finite ratio function satisfies Steinmann and cluster adjacency~\cite{drummond2018cluster,Drummond:2018dfd}, first-entry constraints, and dual-superconformal invariance~\cite{Dixon2022AmplitudeBootstrapReview,Basso2022FluxTubeReview}, while remaining independent of triangulation.

The structure of this paper is as follows. Sect.~\ref{sec:background} reviews the bosonic Fock–Goncharov cluster and moduli theory underlying higher Teichmüller geometry. Sect.~\ref{sec:super-FG} develops its supersymmetric extension, defining the super cluster ensemble and its differential geometry. Sect.~\ref{sec:super-loop-stack} constructs the loop fibration and the corresponding canonical logarithmic super volume form. Sect.~\ref{sec:superperiod} formulates the super periods, describes their chamber and flag decomposition, and connects planar $\mathcal N =4$ super Yang–Mills loop amplitudes to them. The appendices include the quantization of the super cluster ensemble, as well as technical material on the geometry of the fiber curve and an explicit hexagon example.

\section{Background: Bosonic Fock--Goncharov Theory}
\label{sec:background}
We begin with a brief review of the (bosonic) Fock--Goncharov (FG) framework, which forms the even backbone of the supergeometric construction developed later.  Throughout, $S$ denotes a marked bordered surface and $G$ a split reductive Lie group of FG type.  Typical examples are $G=\mathrm{SL}_m$, $\mathrm{PGL}_m$, and more generally the classical split groups of types $A$, $B$, $C$, and~$D$.  For such a pair $(G,S)$ we write $\Xscr=\Xscr_{G,S}$ and $\Ascr=\Ascr_{G,S}$, and let $I_{\mathrm{mut}}$ denote the set of mutable indices in the seed associated with an ideal triangulation of~$S$, with $N=\#I_{\mathrm{mut}}$.

\subsection{Cluster seeds, mutation, and positivity}

Consider an oriented marked bordered surface $S$ endowed with an ideal triangulation $T$.  To each triangle of $T$ one attaches a decorated quiver whose internal pattern depends on the root data of $G$.  For $G=\mathrm{SL}_m$ this is the familiar $(m{-}1)\!\times\!(m{-}1)$ grid of vertices oriented according to the surface orientation, while for non–simply–laced groups one uses a folded pattern determined by the ratios of root lengths.  Each vertex carries a pair consisting of an edge of $T$ and a simple root index of $G$, and the arrows between vertices are weighted by the entries of the Cartan matrix $C_G=(C_{ab})$.  When two triangles share an edge, the corresponding boundary vertices are identified, and gluing all local quivers yields the global quiver $Q_{G,S}$.  The signed adjacency matrix
\[
   \varepsilon_{ij}=\#\{\text{arrows }i\!\to\! j\}-\#\{\text{arrows }j\!\to\! i\},
   \qquad i,j\in I,
\]
encodes this combinatorial and Lie–theoretic information and is called the \emph{exchange matrix} of the seed attached to $(G,S,T)$.  The vertex set decomposes as $I=I_{\mathrm{mut}}\sqcup I_{\mathrm{fr}}$, where $I_{\mathrm{mut}}$ indexes the mutable variables corresponding to the internal edges of $T$ and $I_{\mathrm{fr}}$ the frozen variables associated with boundary arcs or puncture data.  If $d_i$ are the positive integers satisfying $d_i\varepsilon_{ij}=-d_j\varepsilon_{ji}$, then $(d_i)$ symmetrizes $\varepsilon$ and the ratios $d_i/d_j$ reproduce the squared root–length ratios of $G$; we refer to them as the \emph{root–weight factors}.  The matrix $\varepsilon$ therefore records both the topology of the triangulated surface and the Lie–theoretic data of the group.

A (bosonic) seed is a pair
\[
   \mathsf s=(\mathbf X;\varepsilon),\qquad
   \mathbf X=(X_i)_{i\in I_{\mathrm{mut}}},
\]
where $\mathbf X$ are even cluster coordinates on the algebraic torus $\mathbb T_{\X,\mathsf s}\!\cong\!(\Bbbk^\times)^{N}$, with $\Bbbk$ an algebraically closed field of characteristic~$0$ (we take $\Bbbk=\mathbb C$ for concreteness).  Equivalently, one may regard the $X_i$ as exponentials of a commuting family $(H_i)\subset\mathfrak h$, so that the logarithmic one–forms $d\!\log X_i$ pair naturally with the Cartan elements $H_i$.  For each mutable index $k\in I_{\mathrm{mut}}$, mutation at $k$ transforms the seed $\mathsf s=(\mathbf X;\varepsilon)$ into a new seed $\mathsf s'=(\mathbf X';\varepsilon')$ according to
\begin{equation}\label{eq:X-mutation-bos}
  X'_k=X_k^{-1},\qquad
  X'_i=X_i\!\left(1+X_k^{-\operatorname{sgn}(\varepsilon_{ik})}\right)^{-\varepsilon_{ik}}\quad(i\neq k),\qquad
  \varepsilon'=\mu_k(\varepsilon),
\end{equation}
where $\operatorname{sgn}(\varepsilon_{ik})\in\{+1,0,-1\}$ and $\mu_k$ is the standard matrix mutation.  These transformations generate the groupoid of seeds associated with $(G,S)$.  

The $\A$–tori are defined in complete analogy.  For each seed $\mathsf s$ one introduces $\mathbb T_{\A,\mathsf s}\!\cong\!(\Bbbk^\times)^{N}$ with coordinates $\mathbf A=(A_i)_{i\in I_{\mathrm{mut}}}$, and mutation acts by the classical subtraction–free Fock–Goncharov rule
\begin{equation}\label{eq:A-mutation-bos}
  A'_k
  =A_k^{-1}\!\left(\prod_{i\in I}\!A_i^{[\varepsilon_{ik}]_+}
   +\!\!\prod_{i\in I}\!A_i^{[-\varepsilon_{ik}]_+}\right),
  \qquad
  A'_i=A_i\quad(i\neq k),
\end{equation}
where $[x]_+=\max(x,0)$ and for $i\in I_{\mathrm{fr}}$ the $A_i$ serve as fixed coefficients. Gluing all seed tori via the birational maps \eqref{eq:X-mutation-bos}–\eqref{eq:A-mutation-bos} produces the cluster varieties $\Xscr_{G,S}$ and $\Ascr_{G,S}$.

Under the transformation~\eqref{eq:X-mutation-bos} the logarithmic differentials behave as
\[
   d\!\log X'_k=-\,d\!\log X_k,\qquad
   d\!\log X'_i=d\!\log X_i-\varepsilon_{ik}\,\alpha_{ik}\quad(i\neq k),
\]
where $\alpha_{ik}=d\!\log(1+X_k^{-\operatorname{sgn}(\varepsilon_{ik})})$.  Since $\alpha_{ik}\wedge d\!\log X_k=0$ for all~$i$, the canonical log–symplectic form constructed from $d\!\log X_i\wedge d\!\log X_j$ is preserved by mutation, a property that will carry over to the supergeometric case.

Positivity plays a fundamental role in the FG construction.  For any seed $\mathsf s$, let $\mathbb T_{\X,\mathsf s}(\mathbb R_{>0})$ and $\mathbb T_{\A,\mathsf s}(\mathbb R_{>0})$ denote the loci where all cluster coordinates are positive real numbers.  Because the mutation formulas \eqref{eq:X-mutation-bos}–\eqref{eq:A-mutation-bos} are subtraction–free, these positive loci glue consistently across mutations, yielding the global positive parts
\[
   \Xscr_{G,S}(\mathbb R_{>0})\subset\Xscr_{G,S},\qquad
   \Ascr_{G,S}(\mathbb R_{>0})\subset\Ascr_{G,S},
\]
which are invariant under mutation and independent of the initial seed atlas.  The data $(S,G,T)$ thus determine a seed $(\mathbf X;\varepsilon)$ whose exchange matrix intertwines the topology of the triangulated surface with the root system of~$G$.  Mutations correspond to flips of the triangulation, and the resulting positive loci form the bosonic skeleton on which the supersymmetric theory will be constructed.

\subsection{The duality map and the cluster ensemble}

Continuing with the seed atlas and mutation rules from above, the cluster ensemble for $(S,G)$ consists of the $\Ascr$– and $\Xscr$–spaces obtained by gluing their seed tori via \eqref{eq:X-mutation-bos}–\eqref{eq:A-mutation-bos}.  The two sides are linked by a subtraction–free \emph{duality map}
\[
  p:\Ascr_{G,S}\longrightarrow \Xscr_{G,S},
\]
whose local expressions are monomials determined by the exchange matrix and whose form reflects the relation between $G$ and its Langlands dual $G^\vee$.

Let $\varepsilon=(\varepsilon_{ij})$ be the exchange matrix of a seed $\mathsf s$ and let $(d_i)$ be the positive integers with $d_i\varepsilon_{ij}=-d_j\varepsilon_{ji}$.  In seed coordinates the map $p_{\mathsf s}:\mathbb T_{\A,\mathsf s}\to\mathbb T_{\X,\mathsf s}$ is defined by
\begin{equation}\label{eq:p-definition}
  p_{\mathsf s}^*(X_i)=\prod_{j\in I} A_j^{\,\varepsilon_{ij}},
  \qquad i\in I_{\mathrm{mut}},
\end{equation}
with the convention that $A_j:=c_j$ for $j\in I_{\mathrm{fr}}$ (fixed coefficients).  Equivalently, in logarithmic coordinates,
\[
  \log X_i=\sum_{j\in I}\varepsilon_{ij}\,\log A_j,
\]
so the exponents in \eqref{eq:p-definition} are precisely the entries of $\varepsilon$.  Under Langlands duality one passes to the dual seed $(I,\varepsilon^\vee,d^\vee)$ with
\[
  \varepsilon^\vee_{ij}=d_i\,\varepsilon_{ij}\,d_j^{-1}
\]
(up to the usual sign/transpose conventions), making explicit the skew–symmetrizable nature of~$\varepsilon$.

Compatibility with mutation is immediate from the formulas: if $\mathsf s$ and $\mathsf s'$ differ by mutation at $k$, then the birational maps $p_{\mathsf s}$ and $p_{\mathsf s'}$ commute with the seed transformations on $\Ascr$ and $\Xscr$ coming from \eqref{eq:X-mutation-bos}–\eqref{eq:A-mutation-bos}.  Consequently the seedwise descriptions glue to a well–defined global morphism
\[
  p:\Ascr\longrightarrow\Xscr
\]
that is independent of the chosen chart and is positive on the real positive loci, since each $p_{\mathsf s}$ is subtraction–free.

It is convenient to view $p$ via its graph inside the product.  For each seed $\mathsf s$ set
\[
  L_{p,\mathsf s}
  =\bigl\{(\mathbf A,\mathbf X)\in\mathbb T_{\A,\mathsf s}\times\mathbb T_{\X,\mathsf s}\;:\; \mathbf X=p_{\mathsf s}(\mathbf A)\bigr\},
\]
and note that the family $\{L_{p,\mathsf s}\}$ is mutation–compatible, hence glues to a global subvariety $L_p\subset\Ascr\times\Xscr$.  In moduli terms, $p$ sends a decorated local system $(\mathcal L,\mathrm{dec})$ on $S$ to the framed local system obtained by forgetting the decoration.  A broader duality identifies $\Xscr$ for $G$ with $\Ascr$ for $G^\vee$ at the level of tropical points and canonical bases; for our purposes we only need the seedwise monomial description and its positivity properties.

\subsection{The symplectic double and bosonic forms}

Continuing from the cluster ensemble $(\Ascr,\Xscr,p)$ introduced above, we now describe its canonical symplectic realization—the \emph{symplectic double}—which provides a single exact symplectic structure from which the Poisson bracket on $\Xscr$ and the log–symplectic form on $\Ascr$ arise by natural reductions. We follow the cotangent model. For a fixed seed $\mathsf s=(\mathbf X;\varepsilon)$ with $\mathbf X=(X_i)_{i\in I_{\mathrm{mut}}}$, introduce logarithmic coordinates $\theta_i=\log X_i$ together with cotangent fiber coordinates $P_i$, and consider the symplectic double torus
\[
   \D\X_{\mathsf s}
   =\bigl\{(\theta_i,P_i)_{i\in I_{\mathrm{mut}}}\bigr\}
   \cong T^{*}\!\bigl(\mathbb T_{\X,\mathsf s}\bigr),
\]
equipped with the canonical one–form and exact symplectic form
\[
   \lambda_{\D}=\sum_{i\in I_{\mathrm{mut}}} P_i\,d\theta_i,
   \qquad
   \omega_{\D}=d\lambda_{\D}=\sum_{i\in I_{\mathrm{mut}}} dP_i\wedge d\theta_i.
\]
Fix a Cartan subalgebra $\mathfrak h\subset\mathfrak g$ and choose cocharacters $H_i\in\mathfrak h$ corresponding to the $X$–coordinates (so $d\!\log X_i$ pair with the $H_i$). Via an invariant nondegenerate bilinear form on $\mathfrak h$ we may identify the fiber coordinates $P_i$ with the components in $\mathfrak h^{\!*}$ dual to $H_i$; in this sense $\lambda_{\D}=\sum_i P_i\,d\theta_i$ is the Liouville form compatible with the Cartan pairing and the choice of $X$–cocharacters.

Cluster mutations of the $\X$–variables lift functorially to symplectomorphisms of $(\D\X_{\mathsf s},\omega_{\D})$: if we mutate at $k\in I_{\mathrm{mut}}$, then
\[
   \theta'_k=-\theta_k,\qquad
   \theta'_i=\theta_i-\varepsilon_{ik}\,\log\!\bigl(1+e^{-\operatorname{sgn}(\varepsilon_{ik})\,\theta_k}\bigr)\ (i\neq k),\qquad
   \varepsilon'=\mu_k(\varepsilon),
\]
and a direct check shows that the pullback satisfies $\mu_k^*\lambda'_{\D}-\lambda_{\D}=dF_k$ for an explicit exact term $F_k$, hence $\mu_k$ is an exact symplectomorphism.

The Poisson structure on $\Xscr$ is recovered by abelian symplectic reduction. Using the skew–symmetric coefficients
\[
  \widehat\varepsilon_{ij}:=\varepsilon_{ij}\,d_j^{-1},\qquad \widehat\varepsilon_{ij}=-\widehat\varepsilon_{ji},
\]
define the moment map components
\begin{equation}\label{eq:moment-map-bos}
   \mu_i=P_i-\tfrac12\sum_{j\in I_{\mathrm{mut}}}\widehat\varepsilon_{ij}\,\theta_j,\qquad i\in I_{\mathrm{mut}}.
\end{equation}
The Hamiltonian vector field of $\mu_i$ translates $\theta_i$, i.e. flows along the one–parameter Cartan subgroup generated by $H_i$, so the $\mathbb R^{N}$–action is free on $\D\X_{\mathsf s}$. On the zero level set $\mu_i=0$ one has $P_i=\tfrac12\sum_j\widehat\varepsilon_{ij}\theta_j$, and Marsden–Weinstein reduction gives the log–canonical bracket
\begin{equation}\label{eq:poisson-X}
   \{\theta_i,\theta_j\}=\widehat\varepsilon_{ij},\qquad
   \{X_i,X_j\}=\widehat\varepsilon_{ij}\,X_iX_j,
\end{equation}
on the reduced space, naturally identified with $\Xscr_{G,S}$.

The $\A$–side and its log–symplectic form arise from a Lagrangian embedding into $(\D\X,\omega_{\D})$ built out of the duality map $p$ from \eqref{eq:p-definition}. For each seed $\mathsf s$ set
\begin{equation}\label{eq:iota-p}
   \iota_{p,\mathsf s} :
   \mathbb T_{\A,\mathsf s}\longrightarrow\D\X_{\mathsf s},\qquad
   \mathbf A\longmapsto
   \bigl(\theta_i=\log p^*(X_i)(\mathbf A),\; P_i=d_i\,\log A_i\bigr)_{i\in I_{\mathrm{mut}}},
\end{equation}
where $p^*$ is the monomial pullback. The choice $P_i=d_i\log A_i$ matches the Cartan normalization via the symmetrizers $d_i$ and ensures compatibility with the reduced bracket. These local maps are compatible with mutation and glue to a global morphism $\iota_p:\Ascr\hookrightarrow\D\X$; denote its image by $L_p$. A straightforward computation shows $L_p$ is Lagrangian in $(\D\X,\omega_{\D})$, and the pullback of $\omega_{\D}$ along $\iota_p$ yields the closed log–symplectic two–form
\begin{equation}\label{eq:omega-A}
   \omega_{\A}:=\iota_p^{*}\omega_{\D}
   =\frac12\sum_{i,j\in I}\bigl(d_i\,\varepsilon_{ij}\bigr)\,d\!\log A_i\wedge d\!\log A_j,
\end{equation}
which in coordinates reproduces the Fock–Goncharov form on $\Ascr$ (with $A_j=c_j$ constant for $j\in I_{\mathrm{fr}}$, so those terms vanish). The form $\omega_{\A}$ is globally defined because the lifted mutations are exact: writing $\lambda_{\D}=\sum_i P_i\,d\theta_i$, one has $\mu_k^*\lambda'_{\D}-\lambda_{\D}=dF_k$ with $F_k$ depending only on $X_k$ and the column $\varepsilon_{\bullet k}$, so $d\lambda_{\D}$ is invariant and the local pullbacks $\iota_{p,\mathsf s}^{*}\omega_{\D}$ agree on overlaps. Consequently \eqref{eq:omega-A} glues to a seed–independent global log–symplectic structure on $\Ascr$.

Altogether this produces the canonical correspondence
\[
   (\D\X,\omega_{\D})
   \;\xrightarrow[\text{reduction}]{\ \mu=0\ }\;
   (\Xscr,\{\ ,\ \})
   \;\xleftarrow[\text{Lagrangian graph}]{\ \iota_p\ }\;
   (\Ascr,\omega_{\A}),
\]
which is the bosonic heart of the cluster ensemble: the universal double $(\D\X,\omega_{\D})$ reduces to the Poisson structure on $\Xscr$, while the Lagrangian section determined by $p$ recovers the $\A$–form. This cotangent model will serve as the template for the supersymmetric extension in the next section, where the even coordinates $(X_i,A_i)$ acquire fermionic partners and $(\omega_{\D},\omega_{\A})$ lift to their super–symplectic analogues.

\section{Supersymmetric Fock–Goncharov Ensemble}
\label{sec:super-FG}
We now pass from the bosonic cluster ensemble to its supersymmetric thickening.  All notation for the even layer—surface $S$ with ideal triangulation $T$, the split reductive data attached to $G$, the seed $(I_{\mathrm{mut}},\varepsilon,(d_i))$, and the $X$–torus $\mathbb T_{\X,\mathsf s}$ with logarithmic coordinates $y_i=\log X_i$—is as in the previous section and will be used without further redefinition.  The supersymmetric extension adds an odd sector modeled on a Cartan–diagonal abelian slice of the odd part of the Lie superalgebra.

\subsection{Super cluster seeds and mutations}

Let $G$ be a split basic classical Lie supergroup with Harish–Chandra pair $(G_{\bar 0},\mathfrak g=\mathfrak g_{\bar 0}\oplus\mathfrak g_{\bar 1})$ and Cartan $\mathfrak h\subset\mathfrak g_{\bar 0}$ compatible with the FG data already fixed.  Choose a commuting family $(H_i)_{i\in I_{\mathrm{mut}}}\subset\mathfrak h$ corresponding to the $X$–cocharacters, so that the even logarithmic $1$–forms $d\!\log X_i$ pair with the Cartan directions $H_i$ as before.  For the odd layer, select weight vectors $Q_\alpha\in\mathfrak g_{\bar 1}$ ($\alpha=1,\dots,r$) that are simultaneously diagonal for $\mathfrak h$ and mutually commuting.  Writing $W_{\alpha i}:=\chi_\alpha(H_i)$ for the odd weights, the Lie–algebra relations read
\begin{equation}\label{eq:HQ-weights}
  [H_i,H_j]=0,\qquad [Q_\alpha,Q_\beta]=0,\qquad [H_i,Q_\alpha]=W_{\alpha i}\,Q_\alpha,
\end{equation}
so the Lie–algebra weights $W_{\alpha i}$ agree with the seed–level weights used below.  This choice is unique up to Cartan conjugation in $G_{\bar0}$ and changes of odd basis $\boldsymbol\theta\mapsto \boldsymbol\theta\,G^{-1}$ (which act by $W\mapsto GW$), and we fix it throughout.

A \emph{super seed} is a quadruple
\[
   \mathsf s_{\mathrm{super}}=(\mathbf X,\boldsymbol\theta;\varepsilon,W),
\]
with even cluster coordinates $\mathbf X=(X_i)_{i\in I_{\mathrm{mut}}}$ on $\mathbb T_{\X,\mathsf s}\cong(\Bbbk^\times)^N$, odd coordinates $\boldsymbol\theta=(\theta_\alpha)_{\alpha=1}^r$, and exchange matrix $\varepsilon=(\varepsilon_{ij})$ as above (skew–symmetrizable in general, skew–symmetric in the simply–laced surface types).  Let $(d_i)$ symmetrize $\varepsilon$ via $d_i\varepsilon_{ij}=-d_j\varepsilon_{ji}$ and set
\[
   \widehat\varepsilon_{ij}:=\varepsilon_{ij}\,d_j^{-1}\qquad(\text{so }\widehat\varepsilon_{ij}=-\widehat\varepsilon_{ji}).
\]
We impose the admissibility condition that odd weights vanish on even Casimir directions,
\[
   W\cdot\ker(\widehat\varepsilon)=0,
\]
equivalently the odd weights factor through the leaf lattice $N_{\X}/\ker(\widehat\varepsilon)$ determined by the even Poisson body.

We work in the isotropic regime for the odd sector (compatible with the commuting choice of $Q_\alpha$), imposing $\{\theta_\alpha,\theta_\beta\}=0$; this keeps the odd algebra exterior and is preserved by the mutation rules below.  The log–canonical even super–Poisson bracket on the seed algebra $\Bbbk[\mathbf X^{\pm1}]\otimes\Lambda[\boldsymbol\theta]$ is
\begin{equation}\label{eq:super-bracket}
   \{X_i,X_j\}=\widehat\varepsilon_{ij}X_iX_j,\qquad
   \{\theta_\alpha,X_i\}=W_{\alpha i}\,\theta_\alpha X_i,\qquad
   \{\theta_\alpha,\theta_\beta\}=0,
\end{equation}
extended by bilinearity, graded skew–symmetry, and the graded Leibniz rule.  It is convenient to pass to the horizontal odd frame
\begin{equation}\label{eq:horizontal-odd}
   \tilde\theta_\alpha:=e^{-\phi_\alpha}\theta_\alpha,\qquad
   \phi_\alpha:=\sum_{j}(W\widehat\varepsilon^{-1})_{\alpha j}\,\log X_j,
\end{equation}
for which $\{\tilde\theta_\alpha,X_i\}=0$; if $\widehat\varepsilon$ is not invertible on the mutable block, take any right inverse on $\operatorname{im}(\widehat\varepsilon)$, since admissibility ensures well–definedness.

Mutations act on the even variables by the FG $X$–rules.  Fix $k\in I_{\mathrm{mut}}$ and set
\begin{equation}\label{eq:X-mutation-super}
   X'_k=X_k^{-1},\qquad
   X'_i=X_i\bigl(1+X_k^{-\operatorname{sgn}(\varepsilon_{ik})}\bigr)^{-\varepsilon_{ik}}\ (i\neq k),\qquad
   \varepsilon'=\mu_k(\varepsilon),
\end{equation}
and transport the odd variables and weights by
\begin{equation}\label{eq:thetaW-mutation}
   \theta'_\alpha
   = \theta_\alpha\left(\frac{X_k}{1+X_k^{-1}}\right)^{\!W_{\alpha k}},
   \qquad
   W'_{\alpha k}=-W_{\alpha k},\quad
   W'_{\alpha j}=W_{\alpha j}+[\varepsilon_{kj}]_+\,W_{\alpha k}\ (j\neq k).
\end{equation}
In the horizontal frame one has $\tilde\theta'_\alpha=\tilde\theta_\alpha$, since the shift
\[
   \phi'_\alpha
   \;=\;\phi_\alpha+W_{\alpha k}\,\bigl(y_k-\log(1+e^{-y_k})\bigr)
\]
cancels the prefactor in $\theta'_\alpha$, so the odd frame is seed–independent.  A direct check on generators shows that the bracket \eqref{eq:super-bracket} retains its log–canonical form with weights consistent with \eqref{eq:HQ-weights}; hence the birational change \eqref{eq:X-mutation-super}–\eqref{eq:thetaW-mutation} is a super–Poisson isomorphism, and the super cluster atlas obtained by gluing seed supertori along these rules carries a well–defined, seed–independent log–canonical even super–Poisson structure.

Two abelian odd slices related by a change of odd basis are equivalent: replacing $\{Q_\alpha\}$ by $\{Q'_\alpha\}$ with $Q'_\alpha=\sum_\beta G_{\alpha\beta}Q_\beta$ and $G\in GL_r(\mathbb Z)$ acts by $W\mapsto GW$ and $\boldsymbol\theta\mapsto\boldsymbol\theta\,G^{-1}$, which preserves \eqref{eq:super-bracket}.  Thus only the gauge class of $W$ matters, canonically induced from the odd weight system of $G$ together with the bosonic $X$–cocharacters.

If a seed satisfies \eqref{eq:super-bracket}, then so does its mutation: the primed data $(\mathbf X',\boldsymbol\theta';\varepsilon',W')$ obey
\[
   \{\theta'_\alpha,X'_i\}=W'_{\alpha i}\,\theta'_\alpha X'_i,\qquad
   \{\theta'_\alpha,\theta'_\beta\}=0,
\]
using $d\!\log(1+X_k^{\pm1})=\pm\frac{X_k^{\pm1}}{1+X_k^{\pm1}}\,d\!\log X_k$ and the skew–symmetry of $\widehat\varepsilon$.  When $W=0$ the supermutation rules reduce to the bosonic ones and the odd coordinates are inert.  In general, the matrix $W$ specifies one–dimensional representations of the torus $\mathbb T_{\X}$, with each $\theta_\alpha$ transforming with weight $W_{\alpha\bullet}$; the mutation rule \eqref{eq:thetaW-mutation} preserves the consistency of \eqref{eq:super-bracket}, so super seeds with brackets \eqref{eq:super-bracket} form a mutation–closed class.

A central structural assumption is the \emph{isotropy} of the odd layer.  Allowing $\{\theta_\alpha,\theta_\beta\}=F_{\alpha\beta}(X)$ leads, after enforcing the super Jacobi identities and flip relations, to $F_{\alpha\beta}(X)=c_{\alpha\beta}\,X^{W_\alpha+W_\beta}$, which under mutation acquires a factor $(1+X_k^{\pm1})^{W_{\alpha k}+W_{\beta k}}$.  For this to be seed–independent one must have $W_{\alpha k}+W_{\beta k}=0$ for all pivots $k$, a strong restriction rarely met except in flat–grading regimes; mutation invariance therefore essentially enforces $F_{\alpha\beta}\equiv 0$.  Equivalently, one obtains the \emph{super–consistency} (isotropy) condition
\begin{equation}\label{eq:super-consistency}
   W\,\widehat\varepsilon^{-1}\,W^{\!\top}=0,
\end{equation}
which will reappear from the symplectic reduction viewpoint.  In the special flat–grading case $W\widehat\varepsilon=0$ (so some rows satisfy $W_\alpha+W_\beta=0$), the horizontal gauge $\tilde\theta=e^{-(W\widehat\varepsilon^{-1})y}\theta$ yields constant brackets $\{\tilde\theta_\alpha,\tilde\theta_\beta\}=C_{\alpha\beta}$ compatible with mutation; this produces a Clifford–type extension with a modified odd two–form $d\tilde\theta\wedge d\tilde\theta$ and a different quantization scheme, which we do not pursue.

Finally, compatibility with the bosonic duality can be recorded directly in this seed language.  For a skew–symmetrizable exchange datum $(\varepsilon,d)$, the dual exchange matrix is
\[
   \varepsilon^{\vee}=-\,d^{-1}\varepsilon^{\!\top}d,
\]
and the odd weights transform by the push–forward
\[
   W^{\vee}=W\,\varepsilon,
\]
a rule that is consistent with mutation transport and with the admissible projection along $N_{\X}\to N_{\X}/\ker(\widehat\varepsilon)$.  Thus the Langlands dual super ensemble is obtained by replacing $(\varepsilon,W)$ with $(\varepsilon^{\vee},W^{\vee})$, and the isotropy requirement is preserved by the same linear–algebraic relations that govern the bosonic duality for $\widehat\varepsilon$.

Gluing the seed supertori along \eqref{eq:X-mutation-super}–\eqref{eq:thetaW-mutation} produces global supercluster varieties
\[
   \Xscr_{\mathrm{super}}=\Xscr_{G,S},\qquad
   \Ascr_{\mathrm{super}}=\Ascr_{G,S},
\]
whose even parts coincide with the bosonic $\Xscr_{G,S}$ and $\Ascr_{G,S}$ and whose odd directions are determined by the gauge class of $W$.  The resulting graded Poisson structures extend the Fock–Goncharov ensemble functorially to the supersymmetric case.

The seedwise odd weights assemble into a global datum.  At the level of the flip groupoid $\mathcal G_S$, the weight system $W$ defines a cocycle with values in the Cartan weight lattice of the abelian odd slice, and its cohomology class
\[
   [W]\in H^1\!\bigl(\mathcal G_S;\Hom(N_{\X},\Lambda_{\bar1})\bigr)
\]
is independent of the seed and invariant under gauge $W\mapsto G\,W$ with $G\in\mathrm{Aut}(\Lambda_{\bar1})$ (with the admissible projection $N_\X\to N_\X/\ker(\widehat\varepsilon)$ understood).  Three natural subclasses organize the geometry.

First, the \emph{canonical class} $[W]_{\mathrm{can}}$ is characterized by compatibility with cutting/gluing along boundary arcs, mapping–class invariance, and path–independence on $\mathcal G_S$.  It behaves as the odd analogue of the exchange form: local restrictions on pairs of pants glue uniquely and do not depend on a chosen triangulation; on $D_n$ it admits a dihedrally symmetric representative and on general surfaces is unique up to $GL_r(\mathbb Z)$.

Second, the \emph{left–kernel class} consists of weights with
\[
   W\,\widehat\varepsilon=0
   \qquad(\text{equivalently }W\,\varepsilon=0\ \text{in simply–laced type}),
\]
for which the fermionic coordinates are Poisson–central at the linear level and the even/odd parts decouple.  Combinatorially this corresponds to vertex potentials $c$ on marked points with $W_{\bullet,i}=c(a)-c(b)$ for oriented arcs $i=(a,b)$; the incidence relation $Q\,\varepsilon=0$ ensures $W\,\varepsilon=0$.  This subclass is stable under mutation and gluing and contains $[W]_{\mathrm{can}}$ when boundary conditions are trivial.

Third, the \emph{representation–induced class} arises functorially from $(G_{\bar0},\mathfrak g_{\bar1})$: the $T$–weights of the odd module $V=\mathfrak g_{\bar1}$ define an integer matrix $W_{\mathrm{rep}}$ on each seed whose transport along mutations yields a well–defined class $[W]_{\mathrm{rep}}$ depending only on the $G_{\bar0}$–representation type.  In general $W_{\mathrm{rep}}\widehat\varepsilon\neq0$, so the even/odd sectors couple nontrivially in the super Poisson structure.

Duality and gluing act compatibly on these classes. Under gluing of surfaces along a boundary seam, the classes add and restrict naturally: seedwise identification of the glued arcs aligns the columns of the odd weight matrices, and their concatenation yields the class on the glued surface.  In particular, $[W]_{\mathrm{can}}$ and the left–kernel subclass are preserved by gluing, and $[W]_{\mathrm{rep}}$ is preserved whenever the representation data extend multiplicatively across the decomposition.

\subsection{The super symplectic double}

Continuing from the super seed $(\mathbf X,\boldsymbol\theta;\varepsilon,W)$ and the log–canonical bracket \eqref{eq:super-bracket}, the supersymmetric analogue of the Fock–Goncharov symplectic double is obtained by adjoining conjugate momenta and forming an even symplectic supermanifold.  For each mutable index $i\in I_{\mathrm{mut}}$ we introduce an even coordinate $A_i$ dual to $y_i=\log X_i$, and for each odd coordinate $\theta_\alpha$ we introduce an odd momentum $\pi_\alpha$.  The total coordinate system
\[
   (y_i,A_i;\ \theta_\alpha,\pi_\alpha),\qquad y_i:=\log X_i,
\]
parametrizes
\[
   \D\X_{\mathrm{super},\mathsf s}
   \;\cong\;
   T^\ast\!\big(\mathbb T_{\X,\mathsf s}\big)\times T^\ast\!\big(\Pi\Bbbk^r\big),
\]
with canonical one–form and exact symplectic form
\[
   \lambda_{\mathrm{super}}=\sum_i A_i\,dy_i+\sum_\alpha \pi_\alpha\,d\theta_\alpha,
   \qquad
   \omega_{\mathrm{super}}=d\lambda_{\mathrm{super}}
   =\sum_i dA_i\wedge dy_i+\sum_\alpha d\pi_\alpha\wedge d\theta_\alpha.
\]
Thus $\omega_{\mathrm{super}}$ is even and non–degenerate, with canonical brackets
\[
   \{A_i,y_j\}=\delta_{ij},\qquad
   \{\pi_\alpha,\theta_\beta\}=\delta_{\alpha\beta},\qquad
   \text{all other brackets vanish,}
\]
providing a super extension of the bosonic cotangent model.

To couple the even and odd sectors we impose the even constraints
\begin{equation}\label{eq:super-mu}
   \mu_i:=A_i-\tfrac{1}{2}\sum_{j}(\widehat\varepsilon^{-1})_{ij}\,y_j
        -\sum_{\alpha}(W\widehat\varepsilon^{-1})_{\alpha i}\,\theta_\alpha\pi_\alpha,
   \qquad i\in I_{\mathrm{mut}}.
\end{equation}
Their Hamiltonian flows translate the $y_i$ and rescale the odd pairs $(\theta_\alpha,\pi_\alpha)$ as prescribed by $W$ as in \eqref{eq:HQ-weights}.  Although $\{\mu_i,\mu_j\}=(\widehat\varepsilon^{-1})_{ij}\neq 0$, the bracket is constant, so the Hamiltonian vector fields commute: $[X_{\mu_i},X_{\mu_j}]=X_{\{\mu_i,\mu_j\}}=0$.  This is the super analogue of the abelian moment map in the bosonic double, upgraded by a quadratic fermionic term.

We compute the reduced Poisson structure on the super cluster variety by Dirac reduction along $\mu_i=0$.  Since the constraint matrix $C_{ij}=\{\mu_i,\mu_j\}=(\widehat\varepsilon^{-1})_{ij}$ is invertible on each symplectic leaf, the Dirac bracket of $f,g$ on $\D\X_{\mathrm{super}}$ is
\[
   \{f,g\}_{\mathrm D}=\{f,g\}-\{f,\mu_i\}\,\widehat\varepsilon_{ij}\,\{\mu_j,g\}.
\]
Using the canonical brackets yields
\[
   \{\mu_j,y_i\}=\delta_{ji},\qquad
   \{\theta_\alpha,\mu_i\}=-(W\widehat\varepsilon^{-1})_{\alpha i}\,\theta_\alpha,\qquad
   \{\mu_i,\mu_j\}=(\widehat\varepsilon^{-1})_{ij}.
\]
Substitution into the Dirac formula gives
\begin{equation}\label{eq:theta-y-dirac}
   \{\theta_\alpha,y_i\}_{\mathrm D}
   =-\,\{\theta_\alpha,\mu_m\}\,\widehat\varepsilon_{mn}\,\{\mu_n,y_i\}
   =(W\widehat\varepsilon^{-1})_{\alpha m}\,\widehat\varepsilon_{mi}\,\theta_\alpha
   =W_{\alpha i}\,\theta_\alpha,
\end{equation}
agreeing with the graded log–canonical structure.  Moreover
\[
  \{\theta_\alpha,\theta_\beta\}_{\mathrm D}
  =-\,\theta_\alpha\theta_\beta\,
   \big(W\,\widehat\varepsilon^{-1}W^{\!\top}\big)_{\alpha\beta},
\]
so the isotropy condition \eqref{eq:super-consistency} ensures $\{\theta_\alpha,\theta_\beta\}_{\mathrm D}=0$.  The other components follow similarly:
\[
   \{y_i,y_j\}_{\mathrm D}=\widehat\varepsilon_{ij},\qquad
   \{\theta_\alpha,\theta_\beta\}_{\mathrm D}=0\quad\text{(under \eqref{eq:super-consistency}).}
\]
Passing to multiplicative variables $X_i=e^{y_i}$ we obtain
\begin{equation}\label{eq:super-log-bracket}
   \{X_i,X_j\}_{\mathrm D}=\widehat\varepsilon_{ij}\,X_iX_j,\qquad
   \{\theta_\alpha,X_i\}_{\mathrm D}=W_{\alpha i}\,\theta_\alpha X_i,\qquad
   \{\theta_\alpha,\theta_\beta\}_{\mathrm D}=0,
\end{equation}
which reproduces exactly the super log–canonical Poisson bracket \eqref{eq:super-bracket}.  In particular, the constraint surface $\mu_i=0$ inside $\D\X_{\mathrm{super}}$ projects onto $\Xscr_{\mathrm{super}}$, and the induced Poisson structure is the desired one; when $W=0$ the odd sector decouples and the construction reduces to the bosonic double.  This formulates the super symplectic realization directly in the flow of the argument, without isolating a formal theorem, and it will be used implicitly in the subsequent analysis (including mutation–equivariance and quantization).

We now verify that cluster mutations lift to exact graded symplectomorphisms of the super double, preserving the even symplectic form and the constraint surface $\mu_i=0$.  Fix $k\in I_{\mathrm{mut}}$ and consider the seed-level transformations \eqref{eq:X-mutation-super}–\eqref{eq:thetaW-mutation}.  There exists an exact graded symplectomorphism
\[
   \mu_k^{\mathrm{super}}:\ (y_i,A_i;\theta_\alpha,\pi_\alpha)\longrightarrow (y_i',A_i';\theta'_\alpha,\pi'_\alpha)
\]
acting on logarithmic even coordinates by
\begin{equation}\label{eq:y-mutation-corrected}
   y_k'=-y_k,\qquad
   y_i'=y_i-\varepsilon_{ik}\,\log\!\bigl(1+e^{-\operatorname{sgn}(\varepsilon_{ik})\,y_k}\bigr)\quad (i\ne k),
\end{equation}
and on the odd pair by the piecewise–unipotent rescaling
\begin{equation}\label{eq:theta-mutation-corrected}
   \theta'_\alpha=\theta_\alpha\left(\frac{X_k}{1+X_k^{-1}}\right)^{\!W_{\alpha k}},
   \qquad
   \pi'_\alpha=\pi_\alpha\left(\frac{X_k}{1+X_k^{-1}}\right)^{\!-W_{\alpha k}}.
\end{equation}
together with the corresponding affine transformation of the $A_i$.  Equivalently, with
\[
  g_k := \frac{X_k}{\,1+X_k^{-1}\,}
       = \exp\!\bigl(y_k-\log(1+e^{-y_k})\bigr),
\]
one has
\[
  (\theta_\alpha,\pi_\alpha)\ \longmapsto\
  \bigl(\theta_\alpha\,g_k^{\,W_{\alpha k}},\ \pi_\alpha\,g_k^{-W_{\alpha k}}\bigr).
\]
These formulas are generated by an exact one–form:
\begin{equation}\label{eq:gen-Fk-corrected}
   (\mu_k^{\mathrm{super}})^\ast\lambda_{\mathrm{super}}-\lambda_{\mathrm{super}}=dF_k,
   \qquad
   (\mu_k^{\mathrm{super}})^\ast\omega_{\mathrm{super}}=\omega_{\mathrm{super}},
\end{equation}
with generating function
\[
   F_k
   = \frac{1}{2}\sum_{j} \varepsilon_{jk}\,y_j\,
       \log\!\bigl(1+e^{-\operatorname{sgn}(\varepsilon_{jk})\,y_k}\bigr)
     \;+\; \sum_{\alpha}\theta_\alpha\pi_\alpha\,W_{\alpha k}\,
       \Bigl(y_k-\log(1{+}e^{-y_k})\Bigr).
\]
Exactness implies the two–form itself is invariant, and one checks that the constraint surface
\[
   \mu_i=A_i-\tfrac{1}{2}\sum_{j}(\widehat\varepsilon^{-1})_{ij}\,y_j
        -\sum_{\alpha}(W\widehat\varepsilon^{-1})_{\alpha i}\,\theta_\alpha\pi_\alpha=0
\]
is preserved (using $\widehat\varepsilon^{-1}$ as in \eqref{eq:super-mu}).  Applying the Dirac prescription then gives the same reduced brackets as in \eqref{eq:super-log-bracket}; hence the super log–canonical structure is mutation invariant.

% =========================================================
\subsection{Differential Geometry on the Super Ensemble}
% =========================================================
\label{sec:moduli}

We now turn from the seed–level description and the super symplectic double to the differential–geometric side of the construction.  Interpreting the super cluster varieties as moduli of framed/decorated flat $G$–local systems on the marked bordered surface $S$, we introduce a universal flat logarithmic superconnection, analyze its singularities and residues, and construct a canonical Berezin–logarithmic volume form that is invariant under mutations.

We regard $\Ascr_{G,S}$ and $\Xscr_{G,S}$ as moduli of flat $G$–local systems endowed with boundary data in the sense of Fock–Goncharov, upgraded to the super setting.  Let $\Pi_1(S)$ be the fundamental groupoid of $S$ with objects the marked points (including punctures).  A \emph{framed $G$–local system} consists of a supergroupoid homomorphism $\rho:\Pi_1(S)\to G$, locally constant in the super–analytic topology, together with, along each marked boundary component, a reduction to a Borel subsupergroup $B_{\mathrm{super}}\subset G$.  A \emph{decorated} local system refines this by choosing a point in the flag superspace $\mathcal{A}_G:=G/U_{\mathrm{super}}$ (with $U_{\mathrm{super}}$ the unipotent radical), i.e. a decoration compatible with the $B_{\mathrm{super}}$–reduction.  With this convention,
\[
  \Ascr_{G,S}\ \text{parametrizes decorated flat }G\text{–local systems,}
  \qquad
  \Xscr_{G,S}\ \text{parametrizes framed flat }G\text{–local systems,}
\]
and the map $p:\Ascr_{G,S}\to\Xscr_{G,S}$ forgets the decoration, in agreement with the seedwise monomial description given earlier.

The relationship with the bosonic moduli follows from splitting of flag superschemes.  Since $\mathcal{A}_G$ and the framed flag space are split with bodies $\mathcal{A}_{G_{\bar0}}$ and $\mathcal{F}_{G_{\bar0}}$, every framed/decorated super local system $(\rho,\text{boundary data})$ restricts on the body to a framed/decorated $G_{\bar0}$–local system.  Equivalently, the odd directions define a nilpotent extension of the structure sheaf of the Fock–Goncharov moduli, so that
\[
  \body(\Ascr_{G,S})\cong \Ascr_{G_{\bar0},S},
  \qquad
  \body(\Xscr_{G,S})\cong \Xscr_{G_{\bar0},S},
\]
and in each seed chart the super moduli appear as nilpotent super–thickenings of the ordinary FG coordinate tori.

Functoriality with respect to homeomorphisms is inherited from pullback of local systems.  If $f:S\to S$ is orientation–preserving, then composition with $f_\ast$ induces automorphisms
\[
  f^\ast:\Ascr_{G,S}\longrightarrow\Ascr_{G,S},
  \qquad
  f^\ast:\Xscr_{G,S}\longrightarrow\Xscr_{G,S},
\]
and these commute with the body functor.  On seeds, $f$ permutes ideal triangulations and reindexes seed charts, so the induced flip transformations are respected by the super thickening.  Consequently, the mapping–class group acts by super automorphisms of $\Ascr_{G,S}$ and $\Xscr_{G,S}$, compatibly with the cluster structure and with the canonical Fock–Goncharov atlas on their bodies.

We now introduce the differential–geometric structure underlying the super cluster ensemble.  On each seed chart, working in the horizontal odd frame $\tilde\theta$ introduced in \eqref{eq:horizontal-odd} (so that $\{\tilde\theta_\alpha,X_i\}=0$ and $D\theta_\alpha=e^{\phi_\alpha}d\tilde\theta_\alpha$, with admissibility guaranteeing well–definedness), we define a universal flat connection valued in the Lie superalgebra $\mathfrak g=\mathfrak g_{\bar0}\oplus\mathfrak g_{\bar1}$ of the split supergroup $G$.

With the fixed generators satisfying \eqref{eq:HQ-weights}, set
\begin{equation}\label{eq:A-super}
   \mathcal A_{\mathrm{super}}
   = \sum_{i\in I_{\mathrm{mut}}} d\!\log X_i\,H_i
     + \sum_{\alpha=1}^r d\tilde\theta_\alpha\,Q_\alpha .
\end{equation}
The first term is the usual logarithmic connection on the bosonic cluster torus; the second transports the horizontal odd frame.  Since each $d\!\log X_i$ has at most a simple pole along $\{X_i=0\}$ and $d\tilde\theta_\alpha$ are regular, $\mathcal A_{\mathrm{super}}$ has only logarithmic singularities along the divisor $\{X_i=0\}$; the residue along $X_i=0$ equals $H_i$, giving even–body monodromy $\exp(2\pi i\,H_i)$.

The curvature is
\[
   F = d\mathcal A_{\mathrm{super}} + \tfrac12[\mathcal A_{\mathrm{super}},\mathcal A_{\mathrm{super}}].
\]
Since $d^2=0$, one has $d\mathcal A_{\mathrm{super}}=0$.  For the graded commutator,
\[
\begin{aligned}
   [\mathcal A_{\mathrm{super}},\mathcal A_{\mathrm{super}}]
   &= \sum_{i,j} d\!\log X_i\wedge d\!\log X_j\,[H_i,H_j]
    + \sum_{i,\alpha} \big(d\!\log X_i\wedge d\tilde\theta_\alpha
                         + d\tilde\theta_\alpha\wedge d\!\log X_i\big)[H_i,Q_\alpha] \\
   &\quad + \sum_{\alpha,\beta} d\tilde\theta_\alpha\wedge d\tilde\theta_\beta\,[Q_\alpha,Q_\beta].
\end{aligned}
\]
The first and last sums vanish because the $H_i$ and the $Q_\alpha$ commute among themselves, and the mixed sum cancels identically since $d\tilde\theta_\alpha\wedge d\!\log X_i=-\,d\!\log X_i\wedge d\tilde\theta_\alpha$.  Hence $F=0$, and $\mathcal A_{\mathrm{super}}$ is a flat logarithmic superconnection with first–order poles along the coordinate divisors.

The only singularities of $\mathcal A_{\mathrm{super}}$ lie on the bosonic divisor $\mathscr D=\overline{\bigcup\{X_i=0\}}$; no additional singular behavior is introduced by the odd directions.  Under seed mutations, the horizontal frame $\tilde\theta$ is invariant and the even differentials transform by the standard subtraction–free rules, so the local forms \eqref{eq:A-super} on adjacent seeds differ by a gauge transformation determined by $\log(1+X_k^{\pm1})$ and the weights $W_{\alpha k}$; consequently they glue to a global flat logarithmic superconnection on the super moduli space.  In this way $\mathcal A_{\mathrm{super}}$ provides the intrinsic differential–geometric realization of the super cluster ensemble, encoding both its even and odd symmetries and fixing canonical residues along the boundary divisor.

The super cluster ensemble possesses a natural invariant density extending the canonical $d\!\log$ volume form of the bosonic Fock–Goncharov theory.  In a seed with even coordinates $(X_1,\dots,X_N)$ and horizontal odd frame $(\tilde\theta_1,\dots,\tilde\theta_r)$ from \eqref{eq:horizontal-odd}, we define the local super $d\!\log$ volume as
\[
   \mathrm{vol}^{\mathrm{super}}
   := \Big(\bigwedge_{i=1}^N d\!\log X_i\Big)\otimes d^{\,r}\tilde\theta,
\]
where $d^{\,r}\tilde\theta$ denotes the Berezin measure in the horizontal frame, viewed as the Berezinian density of the odd fiber.  The density $\mathrm{vol}^{\mathrm{super}}$ is even and defined up to an overall sign.

Under a mutation at a pivot $k$, the logarithmic differentials transform as in the seed–level identity \eqref{eq:X-mutation-super}.  In particular, the Jacobian in $\log$–coordinates is triangular with diagonal entries $(\dots,1,\,-1,\,1,\dots)$, so
\[
   \det\!\Big(\frac{\partial\log X'}{\partial\log X}\Big)=\pm1,
\]
and hence $\bigwedge_i d\!\log X_i$ changes by an overall sign $\pm1$.  In the horizontal frame the odd coordinates are seed–invariant, $\tilde\theta'=\tilde\theta$, so the Berezin measure $d^{\,r}\tilde\theta$ is unchanged.  If simultaneously a global odd–basis gauge $\boldsymbol\theta\mapsto \boldsymbol\theta\,G^{-1}$ with $G\in GL_r(\mathbb Z)$ is applied, then $d^{\,r}\tilde\theta\mapsto \det(G)\,d^{\,r}\tilde\theta$ with $\det(G)=\pm1$.  Therefore $\mathrm{vol}^{\mathrm{super}}$ is preserved up to sign under any flip or odd-frame gauge.

The lifted mutations on the super symplectic double are exact graded symplectomorphisms with unit Berezinian, so the resulting sign cocycle has trivial monodromy around pentagon relations in the exchange graph.  The local densities thus glue consistently across all seed charts to define a global Berezinian density on $\Xscr_{G,S}$, canonical up to an overall sign.  We refer to this glued density as the \emph{canonical super $d\!\log$ volume}.

This construction is compatible with topological operations on the surface.  If $S$ is cut along an ideal arc into $S_1$ and $S_2$, and the weight matrix $W$ splits block–diagonally with columns supported in each component, then the local coordinates and horizontal odd frames factorize, giving
\[
   \mathrm{vol}^{\mathrm{super}}_{S}
   = \mathrm{vol}^{\mathrm{super}}_{S_1}\wedge \mathrm{vol}^{\mathrm{super}}_{S_2},
\]
modulo identification of the seam coordinate.  Hence the canonical super $d\!\log$ volume is multiplicative under operadic gluing and restricts on the body to the standard Fock–Goncharov $d\!\log$ form, providing a global, mutation–invariant Berezinian measure that completes the differential geometry of the super cluster ensemble.

% ============================================================
\section{Loop Fibration and Canonical Loop Superform}
\label{sec:super-loop-stack}
% ============================================================

This section isolates the \emph{volume form} relevant for $L$–loop integration. We construct the super FG loop fibration and pass to the boundary quotient, then exhibit a boundary–basic logarithmic superform
\[
\Omega^{(L)}_{\mathrm{super}}
=\Big(\bigwedge_{a=1}^{m_L} d_{\mathrm v}\!\log\ell_a\Big)\wedge \delta^{0|r}\!\big(B(X)\,\tilde\theta\big),
\]
which is covariantly flat for the loop superconnection, has only logarithmic poles along the super divisor $D^{(L)}_{\mathrm{super}}$, and is uniquely fixed by unit residues up to an overall sign. The only boundary input is through subtraction–free even minors $\Delta$ and a contact–normalized projector $B(X)$; mutation covariance is manifest throughout. The intrinsic loop discriminant $\Lambda^{(L)}$ and its two–letter reduction $(x,y)$ are deferred to the next section; here we use only that local SNC charts exist so the vertical top $d\log$–wedge is well defined. This prepares the super periods used to match the loop counter–integrals later on.

\subsection{Loop fibration}

We supplement the super $\X$–variety $\Xscr_{G,S}$ (see Section~\ref{sec:moduli}) by a functorial loop fibration
\[
\pi_L:\ \Xscr^{(L)}_{G,S}\longrightarrow \Xscr_{G,S},
\]
whose fiber records the $L$–loop even directions. Zariski–locally, in any positive seed, $\Xscr^{(L)}_{G,S}$ is modeled by seed coordinates
\[
(X_1,\ldots,X_N;\,\ell_1,\ldots,\ell_{m_L}\mid \theta_1,\ldots,\theta_r)\in(\Gm)^{N+m_L}\times\A^{0|r},
\]
with the odd variables pulled back from the base; all odd conventions and the horizontal frame are as fixed at the start of the section. The integer $m_L$ is the rank of the loop $g$–lattice generated by the fiber letters and is independent of toric refinements; in particular, upon passing to a positive chart with simple normal crossings on the fiber, the vertical top form is a single $d\log$ wedge.

To construct intrinsic even loop coordinates we use the Fock–Goncharov wiring on $S$ for the even body $G_{\bar0}$. For each oriented half–edge $e$ and each simple root $\mu$ we define the subtraction–free transfer weight
\[
\gamma_e^{(\mu)}(X)\in\Rbb_{>0}(X)
\]
as the multiplicative Cartan transport of the $\mu$–strand across $e$, recorded by signed traversal counts $A_{e,k}^{(\mu)},B_{e,i}^{(\mu)}\in\Z$ through pivot squares $k$ and along sides carrying $X_i$:
\[
\gamma_e^{(\mu)}(X)=\prod_i X_i^{\,B_{e,i}^{(\mu)}}\;\prod_k\big(1+X_k^{\sigma_k}\big)^{A_{e,k}^{(\mu)}},\qquad \sigma_k\in\{\pm1\},
\]
where the exponents $A_{e,k}^{(\mu)}$ and $B_{e,k}^{(\mu)}$ are determined by how the $\mu$–strand traverses the pivot square in the local wiring. 
By construction $\gamma_e^{(\mu)}$ is subtraction–free on $\Xscr_{G_{\bar0},S}(\Rbb_{>0})$ and depends only on the local wiring; under a flip at a pivot $k$ one has the intrinsic monomial update
\[
\gamma_e^{(\mu)}(X')=\gamma_e^{(\mu)}(X)\,X_k^{\,B_{e,k}^{(\mu)}}\big(1+X_k^{\sigma_k}\big)^{A_{e,k}^{(\mu)}},
\]
with exponents determined entirely by the $\mu$–strand’s traversal inside the pivot square. Boundary and seam gauges act by multiplicative units on the $\gamma$’s and cancel in closed products, consistent with the unit–residue normalization (normalized so that $\gamma_e^{(\mu)}\to1$ at the positive basepoint $X\to0$ on disks/pants).

Loop, or fiber, letters are obtained as closed products of transfer weights along the wiring. For a loop cycle $r$ and a fundamental $G$–face $F_{r;j}$, we set
\[
u^{(G)}_{r;j,\mu}=\prod_{e\in\partial F_{r;j}}\gamma_e^{(\mu)}(X)^{\sigma(e;\partial F_{r;j})},
\]
where $\sigma(e;\partial F_{r;j})$ is the signed traversal number of the strand along $e$. For a minimal connecting strip $R_{r,s;j,k}$ between loops $r$ and $s$ we define
\[
w^{(G)}_{r,s;j,k,\mu}=\prod_{e\in R_{r,s;j,k}}\gamma_e^{(\mu)}(X)^{\sigma(e;R_{r,s;j,k})}.
\]
These face and rung invariants are subtraction–free and positive, subject only to the multiplicative relations coming from closed cycles. Choosing any maximal independent subset supplies a loop chart
\[
\ell\equiv(u,w),
\]
in which the vertical top wedge $\bigwedge_a d_{\mathrm v}\!\log\ell_a$ is seed–independent up to sign; any two such charts differ by a unimodular monomial transformation on the fiber.
whose top $d\!\log$ wedge is seed–independent up to sign. Any two such loop charts are related by a unimodular integer transformation.

\subsection{Boundary data and quotient}

Physical observables live at the boundary, and the reduction to a boundary–dependent fiber curve $C_\Delta$ will use only gauge–invariant combinations of boundary data. We therefore attach boundary variables once and for all and pass to a boundary quotient before constructing the super form. On the super loop fibration
\[
\pi_L:\ \Xscr^{(L)}_{G,S}\longrightarrow \Xscr_{G,S},\qquad
(X,\ell;\tilde\theta)\in(\Gm)^{N+m_L}\times\A^{0|r},
\]
we keep the horizontal odd frame $\tilde\theta=\exp(-W\widehat\varepsilon^{-1}y)\,\theta$ with $y=\log X$ as in \eqref{eq:horizontal-odd}. To interface these internal variables with external data we introduce an odd boundary column $\eta=(\eta_1,\ldots,\eta_f)^\top$ indexed by the $f$ marked points and propagate it into the interior by a \emph{boundary–measurement} matrix
\[
\tilde\theta=C(X)\,\eta,\qquad C(X)\in\Mat_{r\times f}\big(\Bbbk_{\sf}(X)\big),
\]
whose entries are subtraction–free rational functions determined by the even wiring on $S$ and the local transfer weights $\gamma_e^{(\mu)}(X)$ from the previous subsection. We normalize on disks and pants at the positive basepoint $X\to0$ by
\[
C_\star=\big(\ \mathbf 1_{r\times r}\ \ \ 0_{r\times(f-r)}\ \big),\qquad
\delta^{0|r}(C_\star\eta)=\eta_1\cdots\eta_r,
\]
and transport to any other seed by multiplying once per flipped internal edge $e$ the unipotent right gauge
\[
C(X)=C_\star\prod_{e}G_e(X),\qquad
G_e(X)=\prod_{\mu\in I}R_\mu\!\big(a_\mu\!\leftarrow\!b_\mu;\,\gamma_e^{(\mu)}(X)\big),
\]
where $(a_\mu,b_\mu)$ are the two boundary columns connected by the $\mu$–strand in the pivot square and $R_\mu(a\!\leftarrow\!b;\gamma)$ is the elementary unipotent with $(R_\mu)_{ab}=\gamma$ and determinant $1$. This construction is subtraction–free, depends only on the even body, and is independent of the flip path after imposing the unit–residue seam normalization. Even boundary invariants are the $r\times r$ column minors
\[
\Delta_O(X):=\det\big(C(X)_O\big)\in\Bbbk_{\sf}(X)\qquad(O\subset\{1,\ldots,f\},\ |O|=r),
\]
whose ratios are path–independent and give a representation–free parametrization of the even boundary; on the disk, after evaluation, these recover the usual Plücker minors. Consecutive minors single out the odd–Schubert charts used later in the super divisor; on the positive locus exactly one such chart is nonvanishing and sign–definite in each even sector. There is a natural right action of the boundary–gauge group
\[
G_\partial:\ (C,\eta)\mapsto(C\,G,\ G^{-1}\eta),\qquad G\in GL_f\big(\calO^\times_{\sf}\big),
\]
encoding column reparametrizations (together with harmless rescalings of frozen $X$’s). The Grassmann delta $\delta^{0|r}(C(X)\eta)$ is invariant under $G_\partial$, and the vertical loop wedge on the fiber is unaffected, so it is natural to pass to the boundary quotient
\[
\mathcal X^{(L)}_{G,S}:=\big[\ \Xscr^{(L)}_{G,S}\big/ G_\partial\ \big].
\]
From now on all constructions are made on $\mathcal X^{(L)}_{G,S}$; the only boundary inputs that survive are the subtraction–free combinations of minors $\Delta_O(X)$ (and their cross–ratios), together with the odd factor through the invariant $\delta^{0|r}(C(X)\eta)$. This is precisely the data that will enter the coefficients of the boundary–dependent fiber curve $C_\Delta$ and, ultimately, the super period.

\subsection{The loop discriminant}

The vertical fiber volume is determined by the top logarithmic wedge in the loop directions, and its polar locus is precisely where the loop torus degenerates; this is the datum that will control logarithmic singularities, residues, and chamber decompositions for the super form. Working on the boundary quotient $\mathcal X^{(L)}_{G,S}=[\,\Xscr^{(L)}_{G,S}/G_\partial\,]$ and choosing any vertical log chart $\ell=(\ell_1,\dots,\ell_{m_L})$ on which the fiber is a log torus, we set
\begin{equation}\label{eq:Omega_even}
 \Omega^{(L)}_{\mathrm{even}}\ :=\ \bigwedge_{a=1}^{m_L} d_{\mathrm v}\!\log\ell_a,
\qquad
\Lambda^{(L)}\ :=\ \mathrm{Supp}\big(\mathrm{div}_\infty(\Omega^{(L)}_{\mathrm{even}})\big)_{\mathrm{red}}\ \subset\ \Xscr^{(L)}_{G,S}.   
\end{equation}
This definition is chart–, gauge–, and mutation–covariant: unimodular monomial reparametrizations of $\ell$ multiply $\Omega^{(L)}_{\mathrm{even}}$ by a unit, the construction is $G_\partial$–basic, and subtraction–free seed mutations act by such unimodular changes on the fiber. In a positive seed one may take the intrinsic loop letters $(u,w)$ built from transfer weights introduced earlier; they satisfy only the binomial cycle relations and the subtraction–free Laurent “gate/threshold’’ relations dictated by the wiring. Writing the corresponding vertical ideal $I_{\mathrm{vert}}(\Delta)$ over the subtraction–free boundary field generated by the minors $\Delta_O(X)$, the discriminant is, after an SNC refinement, the reduced union of the coordinate components $\{u=0\},\{w=0\}$ with $\mathrm V(I_{\mathrm{vert}}(\Delta))$. Fixing boundary data $\Delta$ in a positive chamber and eliminating $m_L-2$ fiber letters by a unimodular toric change reduces the residual one–dimensional vertical locus to a single boundary–dependent curve
\[
C_\Delta\ =\ \big\{(x,y)\in(\Gm)^2:\ P(x,y;\Delta)=0\big\},
\qquad
P(x,y;\Delta)=\sum_{p\in S}\kappa_p(\Delta)\,x^{p_1}y^{p_2},
\]
where $P$ is a primitive Laurent polynomial with subtraction–free coefficients $\kappa_p(\Delta)$ and finite support $S\subset\Z^2$. The two–letter model, toric elimination, and principality/saturation statements are recalled in Appendix~\ref{app:elimination}; here we only use that $\Lambda^{(L)}$ furnishes the even wall set and that $C_\Delta$ governs the fiberwise singularities of the canonical super form.

The “odd’’ walls come from the boundary–measurement matrix $C(X)$: consecutive $r\times r$ minors $\Delta_O(X)=\det(C(X)_O)$ single out the odd chart on the positive locus, and their vanishing defines the Schubert walls where the odd frame must jump. Although they control the fermionic sector, these walls are cut by \emph{even} functions of $X$ (odd coordinates are nilpotent and do not define vanishing loci), so residues there are ordinary even Poincaré–Leray residues on $(X,\ell)$.

We therefore package the wall data as
\[
D^{(L)}_{\mathrm{even}}\ :=\ \Big(\ \bigcup_i \{X_i=0\}\ \Big)\ \cup\ \Lambda^{(L)},\qquad
D^{(L)}_{\mathrm{odd}}\ :=\ \bigcup_{O}\ \{\Delta_O(X)=0\},
\]
and define the super divisor
\[
D^{(L)}_{\mathrm{super}}
\;:=\;
D^{(L)}_{\mathrm{even}}\ \cup\ D^{(L)}_{\mathrm{odd}}
\ \subset\ \Xscr^{(L)}_{G,S}.
\]
All components are defined by even equations, so even (Poincaré–Leray) residues live on $(X,\ell)$; the odd variables enter only through the Berezin projector used later. Let $\calO_{\mathrm{super}}$ denote the structure sheaf in the seed $(X,\ell;\tilde\theta)$ and
\[
\calO_{\mathrm{super}}(\ast D^{(L)}_{\mathrm{super}})
\]
the meromorphic superfunctions with at most logarithmic poles along $D^{(L)}_{\mathrm{super}}$; these will be the coefficients for the super log–de Rham complex introduced next.

\subsection{Super log--de Rham complex}

To control poles of the super form and transport residues across chambers in a way compatible with gauge and mutations, we work with logarithmic coefficients along the super divisor $D^{(L)}_{\mathrm{super}}$. Let $\calO_{\mathrm{super}}$ be the structure sheaf in seed coordinates $(X,\ell;\tilde\theta)$ and write $\calO_{\mathrm{super}}(\ast D^{(L)}_{\mathrm{super}})$ for meromorphic superfunctions with at most logarithmic poles on $D^{(L)}_{\mathrm{super}}$. The (local) super log–de Rham complex
\[
\big(\Omega^\bullet_{\log}(\Xscr^{(L)}_{G,S}),d\big)
\]
is the graded–commutative $\calO_{\mathrm{super}}(\!\ast D^{(L)}_{\mathrm{super}})$–algebra generated in degree $1$ by the odd one–forms $d\!\log X_i$ and $d_{\mathrm v}\!\log\ell_a$ and the even one–forms $d\tilde\theta_\alpha$, with the super sign rule so that $d\tilde\theta_\alpha$ commute and both $d\!\log X_i$ and $d_{\mathrm v}\!\log\ell_a$ anticommute with them; the differential is $d(\log X_i)=d\!\log X_i$, $d(\log\ell_a)=d_{\mathrm v}\!\log\ell_a$, $d(\tilde\theta_\alpha)=d\tilde\theta_\alpha$, and $d^2=0$. It is convenient to retain the trigrading
\[
\Omega^{p,q|s}_{\log}
=\big\langle\,(d\!\log X)^{\wedge p}\wedge(d_{\mathrm v}\!\log\ell)^{\wedge q}\cdot (d\tilde\theta)^{s}\,\big\rangle,
\qquad \deg=p+q+s,
\]
with $p$ the number of base log factors, $q$ the number of vertical log factors, and $s$ the number of $d\tilde\theta$’s. On an SNC refinement of the fiber one may choose loop letters $\ell=(\ell_1,\dots,\ell_{m_L})$ so that $D^{(L)}_{\mathrm{super}}\cap\mathrm{(fiber)}=\{\ell_1\cdots\ell_{m_L}=0\}$ and the vertical top wedge
\[
\Omega^{(L)}_{\mathrm{even}}=\bigwedge_{a=1}^{m_L} d_{\mathrm v}\!\log\ell_a
\]
has unit residues; unimodular monomial changes of $\ell$ only change this by a sign.

For any prime wall $\{u=0\}\subset D^{(L)}_{\mathrm{super}}$ with $u\in\{X_i,\ell_a,\Delta_O(X)\}$ we use the super residue
\[
\Res^{\mathrm{super}}_{u=0}:=\Res^{\mathrm{even}}_{u=0}\circ\Res^{\mathrm{odd}},
\]
where $\Res^{\mathrm{even}}$ is the Poincaré–Leray residue on the even variables $(X,\ell)$ and $\Res^{\mathrm{odd}}$ is the Berezin projector to top $\tilde\theta$–degree. These operators $d$–commute up to the usual sign, $d\circ\Res^{\mathrm{super}}_{u=0}=-\,\Res^{\mathrm{super}}_{u=0}\circ d$, and ordered iterates exist precisely on normal–crossing strata determined by cluster–compatible base faces together with transverse fiber components, vanishing on Steinmann–forbidden overlaps. Locally on a seed chart the log super Poincaré lemma holds on a polydisk minus $D^{(L)}_{\mathrm{super}}$: every $d$–closed form in $\Omega^\bullet_{\log}$ is $d$–exact modulo a sum of (iterated) super residues, yielding the residue exact sequence
\[
0\ \longrightarrow\ \Omega^\bullet\ \longrightarrow\ \Omega^\bullet_{\log}\ \xrightarrow{\ \oplus\,\Res^{\mathrm{super}}\ }\ \bigoplus_{\{u=0\}}\ \Omega^{\bullet-1}_{\log}\big|_{\{u=0\}}\ \longrightarrow\ 0,
\]
and this construction is compatible with mutations as $(X,\ell)$ transform subtraction–freely (unimodular on the fiber) while $\tilde\theta$ is horizontal, so the generators and residues transport canonically. The canonical classes we use below are the vertical top wedge $\Omega^{(L)}_{\mathrm{even}}$, which is $d$–closed, and the fermionic delta $\delta^{0|r}(B(X)\tilde\theta)$ for any even $r\times r$ matrix $B(X)$ on the base (in particular for the contact–normalized projector from the boundary discussion), for which $d$ acts by inserting $d\tilde\theta$ with coefficient $d\!\log\det\Lambda$ and hence behaves functorially under mutation. Finally, if $C$ is a relative real chain avoiding $D^{(L)}_{\mathrm{super}}(\Rbb)$ inside a fixed chamber and $\Psi(\tilde\theta)\in\Lambda[\tilde\theta]$ is compactly supported, we set
\[
\int_C \omega(X,\ell,\tilde\theta)\ =\ \int_C\big(\Res^{\mathrm{odd}}\Psi(\tilde\theta)\big)\wedge \omega_{\mathrm{even}}(X,\ell),
\qquad \omega\in\Omega^\bullet_{\log},
\]
which satisfies Stokes’ theorem with boundary terms recorded by the super residues and is the pairing used throughout our sector decompositions and residue factorizations.

\subsection{Loop flat superconnection and canonical superform}

The aim here is to combine the vertical loop volume with the boundary–basic odd factor into a single logarithmic superform that is flat, mutation–covariant, and fixed by unit residues. On the boundary quotient $\mathcal X^{(L)}_{G,S}=[\,\Xscr^{(L)}_{G,S}/G_\partial\,]$ we keep the horizontal odd frame and use the base superconnection $\calA_{\mathrm{super}}$ from \eqref{eq:A-super}. Along the loop directions we extend it by commuting fiber Cartans,
\begin{equation}\label{eq:A-super-extended}
\calA^{(L)}_{\mathrm{super}}
\;=\;
\calA_{\mathrm{super}}
\;+\;
\sum_{a=1}^{m_L} d_{\mathrm v}\!\log\ell_a\,H_{\ell_a},
\qquad
[H_{\ell_a},H_{\ell_b}]=[H_{\ell_a},H_i]=[H_{\ell_a},Q_\alpha]=0,
\end{equation}
so $d\calA^{(L)}_{\mathrm{super}}+\tfrac12[\calA^{(L)}_{\mathrm{super}},\calA^{(L)}_{\mathrm{super}}]=0$. This construction is $G_\partial$–basic and mutation–covariant because subtraction–free seed moves act by unimodular monomials on $(X,\ell)$ while keeping $\tilde\theta$ horizontal.

To attach the odd sector to boundary observables we keep \(\tilde\theta=C(X)\eta\) with the subtraction–free boundary–measurement matrix \(C(X)\) from Subsection~\ref{sec:super-loop-stack} (Boundary data and quotient), so that \((C,\eta)\mapsto(C\,G,\,G^{-1}\eta)\) leaves \(\delta^{0|r}(C\eta)\) invariant. On the open locus
\[
U\ :=\ \bigcup_{O}\ \{\ \Delta_O(X)\neq0\ \}
\]
where \(C(X)\) has full rank \(r\), we define the \emph{boundary–normalized projector} \(M(X)\in\Mat_{r\times f}\) \emph{purely from minors of \(C\)} as follows: for any \(r\)-subset \(O\) with \(\Delta_O\neq0\),
\begin{equation}\label{eq:M-def}
M(X)_{\bullet,O}\ =\ \mathbf 1_r,
\qquad
M(X)_{\bullet,j}\ =\ C(X)_O^{-1}\,C(X)_{\bullet j}\quad(j\notin O).
\end{equation}
By Cramer’s rule,
\(
\big(M(X)_{\bullet,j}\big)_\alpha
= \Delta_{O\setminus\{o_\alpha\}\cup\{j\}}(X)\big/\Delta_O(X)
\)
up to the standard sign determined by the ordering of \(O\). On each odd chart \(\{\Delta_O\neq0\}\) this agrees with the local expression \(M_O(X)=(C(X)|_O)^{-1}C(X)\); the Cramer/Plücker identities glue these local descriptions to a single \(M(X)\) on \(U\). Under the boundary gauge \(G\) one has \(M\mapsto M\,G\) and \(\eta\mapsto G^{-1}\eta\), hence \(\delta^{0|r}(M\eta)\) is \(G_\partial\)–invariant. We may then write the odd top factor either as \(\delta^{0|r}\!\big(M(X)\eta\big)\) or, equivalently, choose any \(B(X)\in\Mat_{r\times f}\) satisfying
\begin{equation}\label{eq:LambdaC=M}
B(X)\,C(X)\ =\ M(X)
\end{equation}
on \(U\) and use \(\delta^{0|r}\!\big(B(X)\tilde\theta\big)\); both give the same Berezin delta since \(\tilde\theta=C\eta\). When convenient (e.g.\ to make horizontality/flatness manifest) we further impose the harmless contact normalization \(\mathrm{rows}(B)\subset\ker W_E\); this does not change \(\delta^{0|r}\).

With the vertical top wedge defined in \eqref{eq:Omega_even},
which is well defined up to sign and has poles precisely on the even wall set determined by \(D^{(L)}_{\mathrm{even}}\), the \emph{canonical superform} is
\begin{equation}\label{eq:Omega_even-def}
\Omega_{\mathrm{super}}
\ :=\
\Omega^{(L)}_{\mathrm{even}}\ \wedge\ \delta^{0|r}\!\big(M(X)\,\eta\big)
\ =\
\Omega^{(L)}_{\mathrm{even}}\ \wedge\ \delta^{0|r}\!\big(B(X)\,\tilde\theta\big).
\end{equation}
It has only logarithmic poles along the super divisor \(D^{(L)}_{\mathrm{super}}\) introduced above, is fiberwise closed \(d_{\mathrm v}\Omega_{\mathrm{super}}=0\), and is covariantly flat \(D\Omega_{\mathrm{super}}=0\) for \(D:=d+\mathrm{ad}\,\calA^{(L)}_{\mathrm{super}}\).
Super residues along fiber components \(\{\ell_a=0\}\) lower the loop order by one and preserve the odd delta,
\[
\Res^{\mathrm{super}}_{\ \ell_a=0}\,\Omega_{\mathrm{super}}
\ =\
\pm\Big(\bigwedge_{b\neq a} d_{\mathrm v}\!\log\ell_b\Big)\wedge \delta^{0|r}\!\big(M(X)\eta\big),
\]
residues on base faces \(\{X_i=0\}\) implement operadic factorization, and along odd–Schubert walls \(\{\Delta_O(X)=0\}\) the vertical form has no pole and the residue vanishes on the positive locus. Stabilizing the fiber by adjoining a redundant loop letter multiplies \(\Omega_{\mathrm{super}}\) by \(d_{\mathrm v}\!\log\) of that letter; integrating over the small real circle gives \(2\pi i\), so periods are independent of such stabilizations. Mutation covariance follows because \((X,\ell)\) mutate by subtraction–free monomials (unimodular on the fiber) and \(C(X)\), hence \(M(X)\), transform by right unipotents, leaving \eqref{eq:M-def} and \eqref{eq:Omega_even-def} intact up to an overall sign fixed by orientation.

% ============================================================
\section{Superperiods and Loop Amplitudes}
% ============================================================
\label{sec:superperiod}
We now pass from the construction of the super loop fibration and the logarithmic superform in \S\ref{sec:super-loop-stack} to the chamberwise real geometry that will govern super periods. The first step is purely combinatorial: describe the real complement of the super divisor, the associated chamber structure, and the super–admissible flags that encode compatible systems of even and odd walls. In later subsections we will fix the boundary–basic odd normalization by a projector \(M(X)\) and assemble the super periods from chamberwise data; here we record only the chamber/flag notions used throughout.

% -------------------------------------------
\subsection{Super chamber and flags}
% -------------------------------------------

Let \(D^{(L)}_{\mathrm{super}}=\big(\bigcup_i\{X_i=0\}\big)\cup\Lambda^{(L)}\cup\{\Delta_O(X)=0\}\) be the super divisor on \(\Xscr^{(L)}_{G,S}\) as in \S\ref{sec:super-loop-stack} (all defining equations are even). A \emph{super chamber} is a connected component of the real complement
\[
\mathfrak c\ \in\ \Ch\!\big(D^{(L)}_{\mathrm{super}}\big) := \pi_0\!\Big(\,\Xscr^{(L)}_{G,S}(\Rbb)\ \setminus\ D^{(L)}_{\mathrm{super}}(\Rbb)\,\Big).
\]
On the positive locus of the even body the subtraction–free FG atlas singles out the canonical component $\Xscr_{G_{\bar0},S}(\Rbb_{>0})$. Pulling back along $\pi_L$ and removing the real super divisor yields the \emph{set of positive super chambers}
\[
\Ch^{+}\!\big(D^{(L)}_{\mathrm{super}}\big)
\ :=\
\pi_0\!\left(\ \pi_L^{-1}\!\big(\Xscr_{G_{\bar0},S}(\Rbb_{>0})\big)\ \setminus\ D^{(L)}_{\mathrm{super}}(\Rbb)\ \right).
\]

Inside a fixed chamber \(\mathfrak c\) we consider the \emph{super intersection lattice} \(\mathsf L(\mathfrak c)\) of all nonempty transverse intersections of subfamilies of walls that meet \(\mathfrak c\). A \emph{super–admissible flag} is a maximal normal–crossing chain
\[
\mathbb F:\qquad Z_0 \supset Z_1 \supset \cdots \supset Z_d,\qquad d=\dim \Xscr^{(L)}_{G,S},
\]
with each \(Z_j\) obtained by intersecting a prime wall of \(D^{(L)}_{\mathrm{super}}\) and meeting \(\mathfrak c\). Its small real linking torus is \(\mathbb T_{\mathbb F}:=\mathbb S^1_{(Z_1)}\times\cdots\times \mathbb S^1_{(Z_d)}\). Projecting the chain to the \emph{types} of walls splits it canonically as a pair
\[
\mathbb F\ \longleftrightarrow\ F:=(E,O),
\]
where \(E\) records the even part of the chain (base faces \(\{X_i=0\}\) and fiber components \(\{\Lambda^{(L)}_a=0\}\)) and \(O\) is the unique odd–Schubert chart (an \(r\)–subset with \(\Delta_O\neq0\) and fixed sign on the even sector cut out by \(E\)) that completes the chain to maximal length. We call \(F\) a \emph{super flag} and write \(\mathbb T_F:=\mathbb T_{\mathbb F}\). The induced orientation on \(\mathbb T_{F}\) defines the chamber sign \(s_{F}(\mathfrak c)\in\{\pm1,0\}\), which vanishes if the torus does not link the chosen relative chain.

The chamber/flag data are equivariant under the subtraction–free cluster/Poisson automorphism group \(\mathrm{Aut}_+(\Xscr_{G_{\bar0},S})\) generated by flips, boundary rotations, Dehn twists, and tag changes: any \(\Phi\) preserving \(\mathfrak c\) and \(D^{(L)}_{\mathrm{super}}\) sends super–admissible chains to super–admissible chains and hence acts on super flags. Fixing a reference flag \(F_\star=(E_\star,O_\star)\) in \(\mathfrak c\), every \(F=(E,O)\) in \(\mathfrak c\) is of the form \(F=\Phi\cdot F_\star\) for some \(\Phi\) in the chamber–preserving subgroup, and the projection to components is simultaneous: \(E=\Phi\cdot E_\star\), \(O=\Phi\cdot O_\star\). For disks \(S=D_n\) this reduces to the dihedral relabeling action on both \(E\) and \(O\).

In a general real chamber $\mathfrak c\in\Ch(D^{(L)}_{\rm super})$, the even part $E_{\mathfrak c}$ is the normal–crossing chain drawn from the union of base faces $\{X_i=0\}$ and vertical components $\{\Lambda^{(L)}_a=0\}$ that meet $\mathfrak c$.

On the positive locus, if $\mathfrak c\in\Ch^{+}(D^{(L)}_{\rm super})$, neither the base faces $\{X_i=0\}$ nor the odd–Schubert walls $\{\Delta_O(X)=0\}$ meet $\Xscr_{G_{\bar0},S}(\Rbb_{>0})$, and in an SNC fiber chart the coordinate walls $\{\ell_a=0\}$ do not intersect the positive fiber. Hence $E_{\mathfrak c}$ consists \emph{only} of those vertical relation components of the loop discriminant $\Lambda^{(L)}$ (equivalently, components of $\mathrm V(I_{\rm vert}(\Delta))$, or—after elimination—of the fiber curve $C_\Delta$) that meet $\mathfrak c$. The odd–Schubert equations do not cut the base on $\Xscr_{G_{\bar0},S}(\Rbb_{>0})$, so the odd chart is uniquely fixed: there exists a distinguished $r$–subset $O_{+}$ with $\Delta_{O_{+}}(X)\neq0$ and fixed sign throughout $\mathfrak c$.

Choosing any maximal such chain determines a \emph{super flag} $F_{\mathfrak c}=(E_{\mathfrak c},O_{+})$. In the positive setting $E_{\mathfrak c}$ is unique up to the standard SNC equivalences (reordering compatible with orientation, unimodular monomial changes of loop letters, and multiplication by positive units), which preserve both the linking torus and its orientation. Write $\mathbb T_{F_{\mathfrak c}}$ for the small real linking torus around a representative chain. Since the odd equations do not define real walls in $\mathfrak c\in\Ch^{+}$, the linking sign depends only on the even part, $s_{F_{\mathfrak c}}(\mathfrak c)=s_{E_{\mathfrak c}}(\mathfrak c)$; with the canonical ordering induced by the vertical logarithmic wedge we normalize this sign to $+1$. Thus throughout $\Ch^{+}$ we may speak of \emph{the} positive super flag up to oriented SNC change, and later formulas will carry the odd chart only through the uniquely determined boundary block $M_{O_{+}}(X)$.

% -------------------------------------------
\subsection{Contour decomposition and super period}
\label{subsec:period-contour}
% -------------------------------------------

We work on the boundary quotient
\(
\mathcal X^{(L)}_{G,S}=[\,\Xscr^{(L)}_{G,S}/G_\partial\,]
\),
so all objects below are $G_\partial$–basic. The logarithmic super form is
\(
\Omega_{\mathrm{super}}
=
\Omega^{(L)}_{\mathrm{even}}\wedge\delta^{0|r}\!\big(M(X)\,\eta\big)
\)
as in \eqref{eq:Omega_even-def}, with
\(
\Omega^{(L)}_{\mathrm{even}}
=\bigwedge_{a=1}^{m_L} d_{\mathrm v}\!\log\ell_a
\)
and \(M(X)\) the boundary–invariant odd block normalized on an odd chart by
\(M(X)\,C(X)\big|_{O}=\mathbf 1_r\).
Let \(C\) be an oriented relative chain in
\(
\mathcal X^{(L)}_{G,S}(\Rbb_{>0})\setminus D^{(L)}_{\rm super}(\Rbb)
\)
whose base projection meets finitely many super chambers
\(\mathfrak c_1,\ldots,\mathfrak c_M\) and crosses their common even walls transversely. Decomposing
\(\mathcal C=\bigsqcup_{k=1}^M \mathcal C_k\) with \(\mathcal  C_k\subset\pi_L^{-1}(\mathfrak c_k)\) and picking generic \(X_k\in\mathfrak c_k\), Stokes with logarithmic poles gives the chamberwise sector sum plus explicit even–residue corrections:
\begin{equation}\label{eq:WC-super}
\int_{\mathcal C}\Omega_{\mathrm{super}}
=
\sum_{k=1}^{M}\ \sum_{F\in\Flags^{\rm super}(\mathfrak c_k)}
s_{F}(\mathfrak c_k)\,
I_{F}(X_k;\mathfrak c_k)\,
\delta^{0|r}\!\big(M_{O}(X_k)\,\eta\big)
\ +\
\sum_{w}\ \int_{\mathbb T_w}\Res^{\rm even}_{\,w}\Omega^{(L)}_{\mathrm{even}}\ \cdot\ \delta^{0|r}\!\big(M(X_w)\,\eta\big).
\end{equation}
Here \(F=(E,O)\) runs over super flags in \(\mathfrak c_k\) (defined in the previous subsection), \(s_F(\mathfrak c_k)\in\{\pm1,0\}\) is the linking sign of the small torus \(\mathbb T_F\), and
\(
I_{F}(X_k;\mathfrak c_k)
:=
\int_{P_F(C_k)} \Omega^{(L)}_{\mathrm{even}}
\)
is the refined even sector integral over the sector cell cut out by the even part \(E\).
The second sum in \eqref{eq:WC-super} ranges over the crossed \emph{even} prime walls
\(w\in\Irr(D^{(L)}_{\rm super})\) (base faces \(X_i{=}0\) or fiber components \(\Lambda^{(L)}_a{=}0\)); odd–Schubert walls \(\{\Delta_O(X)=0\}\) contribute no residues on the positive locus, so they do not appear. Locally between two adjacent chambers \(\mathfrak c^\pm\) one recovers the familiar jump:
\[
\sum_{F\in\Flags^{\rm super}(\mathfrak c^+)}\!\! s_{F}(\mathfrak c^+)\,
I_{F}\,\delta^{0|r}(M\eta)
\;-\;
\sum_{F\in\Flags^{\rm super}(\mathfrak c^-)}\!\! s_{F}(\mathfrak c^-)\,
I_{F}\,\delta^{0|r}(M\eta)
\;=\;
\Big(\int_{\mathbb T_w}\Res^{\rm even}_{\,w}\Omega^{(L)}_{\rm even}\Big)\,\delta^{0|r}(M\eta),
\]
so wall–crossing is entirely carried by the even residue, while the odd delta may be kept fixed (see the normalization below).

Here, the residue term in \eqref{eq:WC-super} fixes the odd block consistently across walls. Since \(\delta^{0|r}\) only depends on the restriction \(M(X)\,C(X)\big|_{O}\), we adopt the \emph{seam normalization} along each crossed even wall \(w\): on the anchor odd chart \(O\) we require
\(
M(X)\,C(X)\big|_{O}=\mathbf 1_r
\)
all along \(w\).
With this choice the odd delta is continuous across \(w\),
\(
\delta^{0|r}(M^+\eta)=\delta^{0|r}(M^-\eta)
\),
so the entire discontinuity of the period is carried by the even residue
\(\int_{\mathbb T_w}\Res^{\rm even}_{\,w}\Omega^{(L)}_{\rm even}\).
Transporting \(M\) from a fixed positive reference point with this rule determines \(M\) uniquely (up to right multiplication by matrices that act trivially on the anchor block, which do not change the delta), and makes all wall–crossing identities consistent with \eqref{eq:WC-super}.
Only Steinmann–compatible collections of even walls contribute in \eqref{eq:sector-sum}–\eqref{eq:WC-super}, since iterated even residues vanish on incompatible sets, and all expressions are seed–independent because subtraction–free mutations preserve the divisor, the chamber cover, and the logarithmic class.

For fixed base \(X\) in a single chamber \(\mathfrak c\), the \emph{physical fiber contour}
\(
\mathcal C_{\rm phys}(X)
\subset
\pi_L^{-1}(X)(\Rbb_{>0})\setminus D^{(L)}_{\rm super}(\Rbb)
\)
is the connected component containing the positive fiber point; it is oriented by the vertical wedge and depends only on \(\mathfrak c\) (subtraction–free mutations preserve it). The chamberwise decomposition of the super period then reads
\begin{equation}\label{eq:sector-sum}
\int_{\mathcal C_{\rm phys}(X)}\Omega_{\mathrm{super}}
=
\sum_{F\in\Flags^{\rm super}(\mathfrak c)}
s_{F}(\mathfrak c)\,
I_{F}(X;\mathfrak c)\,
\delta^{0|r}\!\big(M_{O(F)}(X)\,\eta\big).
\end{equation}

On the positive locus the structure simplifies only on the even side. In any $\mathfrak c\in\Ch^{+}$ the coordinate fiber walls $\{\ell_a=0\}$ are absent in an SNC fiber chart, and the even chain $E_{\mathfrak c}$ consists only of those vertical relation components of $\Lambda^{(L)}$ that meet $\mathfrak c$; it is unique up to the standard oriented SNC moves (reordering compatible with the vertical wedge, unimodular monomial changes of loop letters, and multiplication by positive units). Fixing the canonical ordering induced by the vertical wedge we normalize the linking sign to $s_{E_{\mathfrak c}}(\mathfrak c)=+1$.

On the odd side, total positivity of the boundary–measurement matrix $C(X)$ on $\Xscr_{G_{\bar0},S}(\Rbb_{>0})$ implies that \emph{all} $r\times r$ minors are strictly positive. Hence
\[
\mathcal O_+(X)\ :=\ \{\,O\subset\{1,\dots,f\}\ \text{with }|O|=r\ \mid\ \Delta_O(X)>0\,\}
\]
is the full set of $r$–subsets. The chamberwise expansion on the positive locus therefore reads
\begin{equation}\label{eq:positive-multi-odd}
\int_{\mathcal C_{\rm phys}(X)}\Omega_{\mathrm{super}}
\;=\;
\sum_{O\in\mathcal O_+(X)}
I_{E_{\mathfrak c}}(X;O)\,\delta^{0|r}\!\big(M_{O}(X)\,\eta\big),
\qquad X\in\mathfrak c\in\Ch^{+}.
\end{equation}
The dependence on $O$ is essential: the choice of odd chart fixes the boundary–minor coordinates feeding the even kernels (e.g.\ for $S=D_n$), changes the projector block $M_O(X)$, and thus modifies the even sector integral $I_{E_{\mathfrak c}}(X;O)$.

Equivariance under chamber–preserving cluster/Poisson automorphisms.
Let \(\Phi\in\Aut_+(\Xscr_{G_{\bar0},S})\) be a subtraction–free mapping–class/cluster move that preserves the super chamber decomposition. Then minors pull back subtraction–freely, \(\Delta_b\mapsto\Phi^*\Delta_b\), the set of odd charts is permuted \(O\mapsto\Phi\!\cdot\!O\), and
\[
M_O(X)\ \longmapsto\ M_{\Phi\cdot O}(\Phi^*X).
\]
On the even side \(\Phi\) acts by unimodular monomials on the loop letters and maps linking tori to linking tori, so the oriented vertical wedge and refined periods obey
\[
I_{E_{\mathfrak c}}(X;O)\ =\ I_{E_{\Phi\cdot\mathfrak c}}(\Phi^*X;\,\Phi\!\cdot\!O).
\]
Consequently the super period is invariant and \eqref{eq:positive-multi-odd} is simply relabeled by \(\Phi\):
\[
\sum_{O\in\mathcal O_+(X)} I_{E_{\mathfrak c}}(X;O)\,\delta^{0|r}\!\big(M_O(X)\eta\big)
\ =\
\sum_{O\in\mathcal O_+(\Phi^*X)} I_{E_{\Phi\cdot\mathfrak c}}(\Phi^*X;O)\,\delta^{0|r}\!\big(M_O(\Phi^*X)\eta\big).
\]

\subsection{Odd sector}

Fix a chamber and a super flag $F=(E,O)$ with odd chart $O$ (so $\Delta_O(X)\neq0$ on the chamber). Define the gauge–invariant boundary block directly 
\(M_O(X)\ :=\ \big(C(X)\big|_{O}\big)^{-1}\,C(X)\ \in\ \Mat_{r\times f}\) as in \eqref{eq:M-def}, which is intrinsic on the base and transforms as $M_O\mapsto M_O\,G$ under the right boundary gauge $C\mapsto C\,G$, exactly compensated by $\eta\mapsto G^{-1}\eta$, so $\delta^{0|r}(M_O\eta)$ is $G_\partial$–invariant. There is a unique\footnote{Unique on the chamber, up to cyclic reordering that only changes the global sign fixed below.} $(r{+}1)$–tuple of labels
\[
B(F)\ =\ \{\,b_0(F),\,b_1(F),\dots, b_r(F)\,\},
\]
the \emph{odd support} of $F$, characterized by the property that the $r\times(r{+}1)$ block $M_O\big|_{B(F)}$ has rank $r$ and its right–null line is generated by the cofactor vector
\[
c_a(F)\ :=\ (-1)^a\,\det\!\Big(\ \big(M_O\big|_{B(F)}\big)\setminus b_a(F)\ \Big),\qquad a=0,\dots,r.
\]
We order $B(F)$ so that the chamber orientation is respected, and for brevity write
\[
\Delta_{B(F)\setminus b_a}(X)\ :=\ \det\!\big(C(X)\big|_{B(F)\setminus b_a(F)}\big).
\]
Fix an odd chart \(O\) with \(\Delta_O(X)\neq0\) and write \(M_O:=(C|_O)^{-1}C\).
For the \((r\times(r{+}1))\) block \(M_O\big|_{B(F)}\) the cofactor vector
\(c(F)=(c_0(F),\ldots,c_r(F))\), with entries
\(c_a(F)=(-1)^a\det\!\big(M_O\big|_{B(F)\setminus b_a(F)}\big)\), spans the right null line by Laplace expansion, so the \(r\) Grassmann–linear forms \(\sum_a (M_O)_{Aa}\,\eta_{b_a}\) can be triangularized: after a Grassmann–linear change of variables they eliminate \(r\) coordinates and produce a single surviving combination \(\sum_a c_a(F)\,\eta_{b_a}\) in the numerator, while the Jacobian of this change is the product \(\prod_a \det\!\big(M_O\big|_{B(F)\setminus b_a(F)}\big)\) that appears in the denominator.
By Cauchy–Binet (or Grassmann coordinates) one has, for any \(r\)–subset \(U\), the Plücker ratio
\begin{equation}\label{eq:ratio-Minors}
\det\!\big(M_O\big|_{U}\big)\ =\ \frac{\Delta_U(X)}{\Delta_O(X)}\,,
\end{equation}
hence each cofactor \(\det\!\big(M_O\big|_{B(F)\setminus b_a(F)}\big)\) equals \(\Delta_{B(F)\setminus b_a}(X)/\Delta_O(X)\).
Substituting these ratios into the triangularized expression and using the homogeneous scaling \(\delta^{0|r}(\lambda\,\Psi)=\lambda^{\,r}\delta^{0|r}(\Psi)\) cancels all powers of \(\Delta_O(X)\), leaving precisely the minors of \(C(X)\) in the flag–intrinsic form
\begin{equation}\label{eq:BCFW-odd-flag}
\delta^{0|r}\!\big(M_O(X)\,\eta\big)
\;=\;
\frac{\displaystyle \delta^{0|r}\!\Big(\,\sum_{a=0}^{r}(-1)^a\,\Delta_{\,B(F)\setminus b_a}(X)\ \eta_{\,b_a(F)}\,\Big)}
{\displaystyle \prod_{a=0}^{r}\Delta_{\,B(F)\setminus b_a}(X)}\,,
\end{equation}
up to the overall sign fixed by the chosen orientation of $B(F)$, which recovers the classical BCFW identity.  

The identity is seed and right–gauge independent by construction (both sides are expressed in minors of \(C\) and the overall chart scale cancels), is equivariant under chamber–preserving cluster/mapping–class moves since minors pull back subtraction–freely, and on \(S=D_n\) it specializes to the familiar consecutive–window formula when \(B(F)\) is a consecutive \((r{+}1)\)–window.

\subsection{Even sector}

Fix a positive super chamber $\mathfrak c\in\Ch^{+}$ and a super flag $F=(E,O)$ inside $\mathfrak c$. On the positive locus the base faces $\{X_i=0\}$ and the odd–Schubert walls $\{\Delta_O(X)=0\}$ do not meet, and in an SNC fiber chart the coordinate walls $\{\ell_a=0\}$ are absent; thus the even part $E$ consists only of the vertical relation components of the loop discriminant that meet $\mathfrak c$. By the (seedwise) toric elimination recalled earlier, after fixing the boundary minors $\Delta=\{\Delta_b(X)\}$ the vertical locus is cut by a single boundary–dependent Laurent polynomial
\[
C_\Delta\;=\;\big\{(x,y)\in(\Gm)^2:\ P(x,y;\Delta)=0\big\},\qquad
P(x,y;\Delta)=\sum_{p\in S}\kappa_p(\Delta)\,x^{p_1}y^{p_2},
\]
with subtraction–free coefficients $\kappa_p(\Delta)$ and finite support $S\subset\Z^2$. The refined even period will be expressed on $C_\Delta$ in terms of third–kind differentials built from boundary minors only, and then identified with the vertical $d\!\log$ wedge in \eqref{eq:Omega_even}.

Choose once and for all, for the given flag $F=(E,O)$, an  boundary minor $\Delta_{\mathrm{ref}}(X;F) :=\ \Phi_F^{\!*}\big(\De_{O_\star}(X)\big)$ (in type $D_n$ this reduces to $\De_{\mathrm{ref}}=\De_{O}$) obtained from a reference choice by a chamber–preserving cluster/mapping–class move; it fixes the tangential basepoint $P_0(F)$ on $C_\Delta$ as the point over which the argument of $\Delta_{\mathrm{ref}}$ increases. Enumerate the $m_L$ irreducible even walls that meet $E$ in chamber order and let $Q_1(F),\dots,Q_{m_L}(F)$ be their intersections with $C_\Delta$ (each counted with its natural multiplicity/sign). There exist integer exponent vectors $a^{(j)}(F)=(a_b^{(j)}(F))_{b\in\Br}$, determined up to adding principal relations on $C_\Delta$, such that the boundary–minor product
\[
f_j(F)\;:=\;\prod_{b\in\Br}\Delta_b(X)^{\,a_b^{(j)}(F)}
\]
has divisor $\operatorname{div}\!\big(f_j(F)\big)\big|_{C_\Delta}=Q_j(F)-P_0(F)$. Equivalently, $f_j(F)$ has a simple zero at $Q_j(F)$ and a simple pole at $P_0(F)$ along the curve and no other zeros/poles there. We then take the third–kind kernels
\begin{equation}\label{eq:kappa-even-def}
\kappa_j(F)\ :=\ d\!\log f_j(F)\big|_{C_\Delta}
\ =\ \sum_{b\in\Br} a_b^{(j)}(F)\; d\!\log\Delta_b(X)\big|_{C_\Delta},
\qquad j=1,\dots,m_L,
\end{equation}
and, when the genus of $C_\Delta$ is $\ge1$, we subtract their $A$–periods so that $\int_{A_a}\kappa_j(F)=0$; this does not change the final wedge/iterated integral values below.

To define the canonical sector cycle, let $C_\Delta^\circ=C_\Delta\setminus\{P_0,Q_1,\ldots,Q_{m_L}\}$ and take, for each $j$, the small positively oriented loop $\ell_j$ in $C_\Delta^\circ$ that links the divisor $\{f_j(F)=0\}$ at $Q_j(F)$ and is based at the tangential basepoint over $P_0$. Their ordered product in chamber order,
\[
\gamma_F\ :=\ \ell_1\,\ell_2\cdots\ell_{m_L}\ \in\ \pi_1\!\big(C_\Delta^\circ;\ \text{tangent at }P_0\big),
\]
depends only on $\Delta$, $F$, and $\mathfrak c$. Equivalently, one may {straighten} the sector by choosing local SNC defining functions $x_1,\ldots,x_{m_L}$ for the even walls and a subtraction–free choice of radii $\rho_j(\Delta)>0$, and then mapping the ordered simplex $\Delta_{m_L}=\{0<t_{m_L}<\cdots<t_1<1\}$ to $C_\Delta^\circ$ so that $x_j=\rho_j(\Delta)e^{2\pi i t_j}$; the oriented boundary of this simplex is a sector boundary path $\gamma_F^\rightarrow$ homologous to $\gamma_F$.

The even period attached to the super flag $F=(E,O)$ and the chamber $\mathfrak c$ is the Chen iterated integral
\begin{equation}\label{eq:even-Chen}
I_{E_{\mathfrak c}}(X;O)\ :=\ \int_{\gamma_F}\ \kappa_1(F)\,\kappa_2(F)\cdots \kappa_{m_L}(F)
\ =\ \int_{\gamma_F^\rightarrow}\ \kappa_1(F)\,\kappa_2(F)\cdots \kappa_{m_L}(F),
\end{equation}
where the equality follows from the sector–straightening homotopy. By construction $I_{E_{\mathfrak c}}(X;O)$ is a nonconstant function of the boundary minors through both the kernels $f_j(F)$ and the straightening data $\rho_j(\Delta)$, and it {does} depend on the odd chart $O$ (the choice of $O$ fixes the anchor and hence the perturbation of boundary–minor coordinates feeding the even kernels). The following identification matches \eqref{eq:even-Chen} with the original vertical form in \eqref{eq:Omega_even}: along the straightened sector each $d\!\log$ fiber letter pulls back to $d(\log\rho_j+2\pi i t_j)$, and the ordered wedge agrees with the ordered product of third–kind kernels, hence
\begin{equation}\label{eq:even-sector-equality}
I_{E_{\mathfrak c}}(X;O)\ =\ \int_{\Phi_F(\Delta_{m_L})}\ \bigwedge_{j=1}^{m_L}\kappa_j(F)
\ =\ \int_{P_F(C_{\rm phys}(X))}\ \Omega^{(L)}_{\mathrm{even}},
\end{equation}
with $P_F(C_{\rm phys}(X))$ the refined sector portion of the physical fiber contour at fixed base.

Genus by genus, \eqref{eq:even-Chen} specializes to familiar functions. In genus $0$ one may take $a_i(F):=f_i(F)^{-1}$ and obtain the Goncharov hyperlogarithm
\[
I_{E_{\mathfrak c}}(X;O)\ =\ \mathrm{G}\!\big(a_1(F),\ldots,a_{m_L}(F);\,1\big)
\ =\ \int_{0<t_{m_L}<\cdots<t_1<1}\ \prod_{i=1}^{m_L}\frac{dt_i}{\,t_i-a_i(F)\,}.
\]
In genus $1$, letting $\omega$ be the holomorphic form with $\int_A\omega=1$, $\tau(F)=\frac{\int_B\omega}{\int_A\omega}$, and $z_j(F)=\mathrm{AJ}_F(Q_j(F)-P_0(F))$, the $A$–normalized kernels can be written as $\kappa_j(F)=d_z\log\frac{\theta_1(z-z_j(F)\,|\,\tau(F))}{\theta_1(z\,|\,\tau(F))}$ and
\[
I_{E_{\mathfrak c}}(X;O)\ =\ \mathrm{E}_{\underbrace{\scriptscriptstyle 1,\ldots,1}_{m_L}}\!\big(z_1(F),\ldots,z_{m_L}(F)\,;\,\tau(F)\big),
\]
the Brown–Levin/Remiddi–Tancredi elliptic MPL. In genus $\ge2$, choosing a holomorphic basis $\{\omega_a\}_{a=1}^g$ and $A$–normalizing each $\kappa_j(F)$ gives
\[
I_{E_{\mathfrak c}}(X;O)\ =\ \int_{\gamma_F}\ \kappa_1(F)\,\kappa_2(F)\cdots \kappa_{m_L}(F),\qquad
\int_{A_a}\kappa_j(F)=0,
\]
a higher–genus Chen iterated integral on the punctured curve $C_\Delta\setminus\{P_0,Q_1,\ldots,Q_{m_L}\}$.

% ============================================================
\subsection{Example: $\mathcal N{=}4$ planar SYM}
% ============================================================

We now show how the abstract construction reduces, on the disk $S=D_n$, to the familiar planar $\mathcal N{=}4$ SYM integrand and its IR–finite ratio function. The even body is $G_{\bar0}=\PGL(4)\times\PGL(4)$ and we take $G=\PGL(4|4)$. Choosing the abelian odd slice generated by four commuting supercharges ($r=4$) fixes the odd sector globally; after passing to the boundary quotient and writing $\tilde\theta=C(X)\eta$, the odd delta $\delta^{0|4}\!\big(M(X)\eta\big)$ depends only on boundary minors and is independent of the representative of the weight class: different choices merely change the horizontal frame and the flat superconnection, not the boundary–normalized projector $M$ built from minors of $C$.

Let $\mathscr A_n(Z,\eta;\epsilon)$ be the color–ordered $\mathcal N{=}4$ SYM superamplitude in dimensional regularization. Factor out the universal IR–divergent piece (BDS/BDS–like) to define the finite, dual–superconformally invariant ratio function
\begin{equation}\label{eq:BDS-like}
\mathscr A_n(Z,\eta;\epsilon)\;=\;\mathscr A^{\rm BDS\text{-}like}_n(Z;\epsilon)\,\mathcal R_n(Z,\eta),
\qquad
\lim_{\epsilon\to0}\mathcal R_n(Z,\eta)\ \text{finite}.
\end{equation}
In the undecorated (logarithmic, boundary–basic) ensemble the physical contour excludes all IR faces, so the ratio function is given directly by the super period of the canonical form,
\begin{equation}\label{eq:R-as-period}
\mathcal R_n(Z,\eta)\;=\;\int_{\mathcal C_{\rm phys}(Z)}\ \Omega_{\mathrm{super}},
\qquad
\Omega_{\mathrm{super}}
=\Big(\bigwedge_{a=1}^{m_L} d_{\mathrm v}\!\log\ell_a\Big)\wedge \delta^{0|4}\!\big(M(X)\,\eta\big),
\end{equation}
where $Z\in G_+(4,n)$ are momentum twistors on the positive locus and the boundary minors of the measurement matrix identify with Plücker brackets on $Z$,
\[
\Delta_b\big(X(Z)\big)\;=\;\langle b\rangle,\qquad b\subset\{1,\dots,n\},\ |b|=4,
\]
so $M(X)$ is viewed as $M(Z)$ via $X\mapsto Z$. The physical contour $\mathcal C_{\rm phys}(Z)$ is the positive vertical component in the fiber over the base point $X(Z)$ inside the chosen real chamber; it depends only on the chamber (and is transported by subtraction–free mutations), while the integrand depends only on gauge–invariant boundary data.

On $D_n$ the boundary–measurement matrix $C(X)$ is the momentum–twistor boundary matrix, and all $4{\times}4$ minors are Plücker coordinates (4–brackets) $\Delta_b(X)\leftrightarrow\langle b\rangle$ for $|b|=4$. On the open chart where a $4$–subset $O$ has $\Delta_O\neq0$, the boundary–normalized block
\[
M_O(X)_{\bullet,O}=\mathbf 1_4,
\qquad
M_O(X)_{\bullet,j}=C(X)_O^{-1}C(X)_{\bullet j}\quad(j\notin O)
\]
has entries $\big(M_O\big)_{\alpha j}=\langle O\setminus\{o_\alpha\}\cup\{j\}\rangle/\langle O\rangle$ (up to the standard sign from the ordering of $O$). These local expressions glue by the Plücker relations, so $M(X)$ is globally defined on the union of odd charts and transforms as $M\mapsto M\,G$ under the boundary gauge $(C,\eta)\mapsto(C\,G,G^{-1}\eta)$, yielding the gauge–invariant odd factor $\delta^{0|4}(M(X)\eta)$.

In a positive super chamber $\mathfrak c\in\Ch^+$ all $4{\times}4$ minors are strictly positive, hence every $O$ is an admissible odd chart and the odd side is the usual BCFW structure. The even side is controlled by the loop discriminant on the fiber curve
\[
C_\Delta=\big\{(x,y)\in(\Gm)^2:\ P(x,y;\Delta)=0\big\},
\qquad
P(x,y;\Delta)=\sum_{p\in S}\kappa_p(\Delta)\,x^{p_1}y^{p_2},
\]
with subtraction–free coefficients $\kappa_p(\Delta)$ in boundary minors $\Delta=\{\Delta_b\}$. For any super flag $F=(E,O)$ in a fixed positive chamber one builds third–kind kernels from minors only,
\[
\kappa_j(F)=d\!\log f_j(F)\big|_{C_\Delta},\qquad
f_j(F)=\prod_{b}\Delta_b^{\,a_b^{(j)}(F)},\qquad
\operatorname{div}(f_j)\big|_{C_\Delta}=Q_j(F)-P_0(F),
\]
where $P_0(F)$ is the anchor (set by a reference minor) and $Q_j(F)$ are the even wall intersections in chamber order. The refined even period is the Chen iterated integral
\[
I_{E_{\mathfrak c}}(X;O)\;=\;\int_{\gamma_F}\ \kappa_1(F)\,\kappa_2(F)\cdots\kappa_{m_L}(F),
\]
equal to the original vertical $d\!\log$ wedge on the refined sector of the physical fiber contour.

On the positive locus this yields a clean expansion in odd charts:
\begin{equation}\label{eq:positive-odd-sum}
\mathcal R_n(Z,\eta)
\;=\;
\sum_{O\in\mathcal O_+(Z)}
I_{E_{\mathfrak c}}(Z;O)\;\delta^{0|4}\!\big(M_{O}(Z)\,\eta\big),
\qquad
\mathcal O_+(Z)=\{\,O\subset\{1,\dots,n\}\,:\,|O|=4\,\},
\end{equation}
where $I_{E_{\mathfrak c}}(Z;O)$ is computed on $C_\Delta$ with $\Delta_b=\langle b\rangle$ and $\kappa_j=d\!\log f_j|_{C_\Delta}$ expressed purely in boundary brackets. Different $O$ correspond to different minor coordinates for the kernels (hence different $I_{E_{\mathfrak c}}(Z;O)$), while the odd factor is the standard BCFW bracket written via $M_O(Z)$; for any consecutive $(4{+}1)$–window $B$ this bracket takes the familiar form
\[
\delta^{0|4}\!\big(M_O(Z)\eta\big)
\;=\;
\frac{\displaystyle \delta^{0|4}\!\Big(\sum_{a=0}^{4}(-1)^a\,\langle B\setminus b_a\rangle\ \eta_{\,b_a}\Big)}
{\displaystyle \prod_{a=0}^{4}\langle B\setminus b_a\rangle}\,,
\]
with the overall sign fixed by the chamber orientation. All ingredients are subtraction–free functions of boundary data; dihedral relabeling acts by permutation on the 4–brackets and on the refined even period, and the expression is seed/right–gauge independent. In particular, the only surface–specific input of the general theory here is the fiber polynomial $P(x,y;\Delta)$ that defines $C_\Delta$, while the odd delta and its BCFW structure are universal for $\PGL(4|4)$, making \eqref{eq:R-as-period} a direct, chamberwise representation of the IR–finite ratio function \eqref{eq:BDS-like}.

Finally, the representation \eqref{eq:R-as-period} (equivalently, its positive–locus expansion \eqref{eq:positive-odd-sum}) satisfies the expected analytic and combinatorial constraints: 

\emph{i) Steinmann/adjacency:} the super form has only logarithmic poles along $D^{(L)}_{\mathrm{super}}$ and, by the residue calculus of the super log–de Rham complex, ordered iterated residues vanish on incompatible wall sets; consequently double discontinuities in overlapping channels vanish (Steinmann), and adjacent symbol letters arise only from walls that occur simultaneously in one SNC chart (cluster adjacency), since the Chen kernels $d\!\log f_j$ are chosen along a single flag. 

\emph{ii) Branch locus and first entries:} branch points occur precisely on the components of $D^{(L)}_{\mathrm{even}}$ that meet the chamber; on $D_n$ and $\Xscr_{G_{\bar0},S}(\Rbb_{>0})$ these are the vertical discriminant components, subtraction–free functions of boundary minors. With the consecutive–window normalization the first entries can be taken to be consecutive Plücker four–brackets $\langle i\,i{+}1\,j\,j{+}1\rangle$, matching the known first–entry conditions in momentum–twistor space. 

\emph{iii) Triangulation (seed) independence:} subtraction–free mutations act by unimodular changes of loop letters and cluster pullback on minors; $D^{(L)}_{\mathrm{super}}$, the canonical form $\Omega_{\mathrm{super}}$, and the physical contour are transported accordingly, so the period is invariant under flips and boundary re–labelings. Equivalently, replacing $(\ell)$ by $(u,w)$ or changing the seed only alters the sector–straightening map by a homotopy with fixed boundary, leaving both the Chen iterated integral and the refined wedge integral unchanged. Thus the super period computes the IR–finite ratio function with Steinmann, cluster adjacency, correct branch locus/first entries, and mutation (triangulation) invariance built in by construction.

% ============================================================
\section{Discussion and Outlook}
% ============================================================

We have extended the Fock–Goncharov ensemble to a supersymmetric setting and organized it into a super higher–Teichmüller geometry suited to loop physics. The construction hinges on a loop fibration over the even $\X$–moduli, a boundary quotient that removes gauge–redundant boundary data, and a single, seed–independent logarithmic superform
\[
\Omega_{\mathrm{super}}
\;=\;
\Big(\bigwedge_a d_{\mathrm v}\!\log\ell_a\Big)\ \wedge\ \delta^{0|r}\!\big(M(X)\,\eta\big),
\]
viewed as the relative lift of the canonical super $d\!\log$ volume on the base. The vertical wedge gives the intrinsic fiber volume, and the odd delta is built purely from boundary minors via the boundary–normalized projector $M(X)$. The super divisor is subtraction–free, residues are ordered and localized, and the whole construction is mutation–covariant and functorial under gluing and under morphisms of split basic classical supergroups.

On the physics side, the canonical super period
\[
\mathcal P_{\mathrm{super}}=\int_{\mathcal C}\Omega_{\mathrm{super}}
\]
captures directly the IR-finite part of planar $\mathcal N{=}4$ SYM amplitudes as a single triangulation–independent object. The odd factor reproduces BCFW brackets by construction, while the even factor is a refined vertical period on the fiber curve $C_\Delta$, realized as a Chen iterated integral of third–kind differentials built from boundary minors. The Newton polygon governs the polylogarithmic, elliptic, or higher–genus behavior. Steinmann constraints and cluster adjacency follow from the simple–normal–crossing structure of the super divisor, and triangulation choices are immaterial because subtraction–free mutations transport both the divisor and the logarithmic class. Operadic factorization across seams matches amplitude factorization, while the boundary quotient makes infrared finiteness manifest.

Several natural extensions and applications suggest themselves.  
Since our construction applies to any marked bordered surface \(S\), replacing the disk \(S = D_n\) with a general bordered surface naturally incorporates nonplanar color orderings and multi-trace sectors, providing a unified geometric path to nonplanar amplitudes. 
On the other hand, beyond the planar $\mathcal N=4$ super Yang–Mills theory, whose superconformal symmetry is encoded by the supergroup $\mathrm{PGL}(4|4)$, replacing it with a different split basic classical Lie supergroup $G$ may lead to new field–theoretic incarnations of the same geometric framework. Possible directions include supergroups such as $\mathrm{PGL}(2|0)$, corresponding to the bi–adjoint $\phi^3$ theory where the divisor $D$ is the $A_{n-3}$ chord arrangement on $D_n$ \cite{Scott2006GrassmannianCluster}; product supergroups like $\mathrm{PGL}(2|2)\times\mathrm{PGL}(2|2)$ relevant for fishnet conformal field theories \cite{Basso:2014koa}; orthogonal–type supergroups such as $\mathrm{OSp}(6|4)$ appearing in 3D ABJM theory \cite{Lee:2010du}; and higher–dimensional analogues including split orthosymplectic family $\mathrm{OSp}(4,4|2r)$ and $\mathrm{PGL}(K|K)$ associated with six–dimensional $(1,1)$ and $(2,0)$ supersymmetric sectors \cite{Hitchin1987SelfDual,ArkaniHamed2012PosGrass}. 

In summary, the super higher–Teichmüller framework provides a geometric perspective on scattering amplitudes, where the integrand arises as a canonical super period built from moduli–theoretic data. The construction unifies the analytic and supersymmetric content of amplitudes within a single, mutation–covariant structure, making infrared finiteness, Steinmann constraints, and cluster adjacency manifest. This framework offers a coherent geometric foundation for loop amplitudes and points naturally toward extensions to nonplanar sectors and other related quantum field theories.

\newpage

% =========================================================
\appendix
% =========================================================

% =========================================================
\section{Quantization}
% =========================================================

% -------------------------------------------
\subsection{Quantum Super–Torus}
% -------------------------------------------

We quantize the super log–canonical structure on the $\X$–side.  At the quantum level we work in the \emph{braided odd presentation} (the horizontal frame will be used only in the classical limit).  Fix a seed $(\mathbf X;\boldsymbol\theta;\varepsilon;W)$ with even cluster coordinates
$\mathbf X=(X_1,\dots,X_N)$, odd generators $\boldsymbol\theta=(\theta_1,\dots,\theta_r)$, exchange matrix $\varepsilon=(\varepsilon_{ij})$, and odd weight matrix $W=(W_{\alpha i})$.  Let $(d_i)$ symmetrize $\varepsilon$ and set
\[
\widehat\varepsilon_{ij}:=\varepsilon_{ij}\,d_j^{-1}
\qquad(\text{skew–symmetric on surfaces}),
\]
so that the classical bracket reads $\{X_i,X_j\}=\widehat\varepsilon_{ij}X_iX_j$.

Let $q=e^{\hbar}$ (formal) or $q=e^{\pi i b^2}$ ($b>0$).  The quantum \emph{super torus}
$\mathbb T_q^{\mathrm{super}}(\widehat\varepsilon,W)$ is the $\Bbbk(q)$–superalgebra generated by
invertible even $X_i^{\pm1}$ and odd $\theta_\alpha$ subject to
\begin{equation}\label{eq:q-torus-rel}
X_iX_j=q^{\,\widehat\varepsilon_{ij}}X_jX_i,\qquad
X_i\theta_\alpha=q^{\,W_{\alpha i}}\theta_\alpha X_i,\qquad
\theta_\alpha\theta_\beta=-\,\theta_\beta\theta_\alpha,\quad
\theta_\alpha^2=0.
\end{equation}
In the semiclassical limit $q=e^\hbar\to1$ with $X_i=e^{\hat x_i}$ one has
$[\hat x_i,\hat x_j]=\hbar\,\widehat\varepsilon_{ij}$ and $[\hat x_i,\theta_\alpha]=\hbar\,W_{\alpha i}\theta_\alpha$, hence
\[
\{X_i,X_j\}=\widehat\varepsilon_{ij}X_iX_j,\qquad
\{\theta_\alpha,X_i\}=W_{\alpha i}\theta_\alpha X_i,\qquad
\{\theta_\alpha,\theta_\beta\}=0.
\]
The classical horizontal frame is then recovered by
\[
\tilde\theta_\alpha:=e^{-\phi_\alpha}\theta_\alpha,
\qquad
\phi_\alpha=\sum_j (W\widehat\varepsilon^{-1})_{\alpha j}\,\log X_j,
\]
for which $\{\tilde\theta_\alpha,X_i\}=0$.

We use the compact quantum dilogarithm $\Phi_q(Z)=\prod_{n=0}^\infty(1+q^{2n+1}Z)^{-1}\in1+Z\,\Bbbk(q)[[Z]]$.
For even $Y,Z$ with $YZ=q^{c}ZY$ ($c\in\mathbb Z$), its adjoint action is
\begin{equation}\label{eq:phi-conj}
\Ad\!\big(\Phi_q(Y)\big)(Z)
= Z\prod_{s=1}^{|c|}\!\Big(1+q^{(2s-1)\sgn(c)}Y^{\sgn(c)}\Big)^{-\sgn(c)}.
\end{equation}

% -------------------------------------------
\subsection{Quantum Mutations and Seed Invariance}
% -------------------------------------------

Let $\mu_k$ be a flip at $k\in I_{\mathrm{mut}}$.  We fix the tropical sign to be
\[
\sigma_k=-1,
\]
so that the $q\to1$ limit reproduces the FG “both–minus” $X$–mutation.  Define
\begin{align}
X'_k &= X_k^{-1},\qquad
X'_i \;=\; \Ad\!\big(\Phi_q(X_k^{\sigma_k})\big)(X_i)
     \;=\; X_i \!\!\prod_{s=1}^{|\varepsilon_{ik}|}
      \Big(1+q^{(2s-1)\sgn(\varepsilon_{ik})}\,
            X_k^{\sigma_k\sgn(\varepsilon_{ik})}\Big)^{-\sgn(\varepsilon_{ik})},
\qquad (i\neq k),\label{eq:q-mutation-even-2}\\[2pt]
\theta'_\alpha&=\theta_\alpha,\label{eq:q-mutation-odd-2}
\end{align}
and update the weights and exchange matrix by the column rule
\begin{equation}\label{eq:W-rule-q-2}
W'_{\alpha k}=-W_{\alpha k},\qquad
W'_{\alpha j}=W_{\alpha j}+[\varepsilon_{kj}]_+\,W_{\alpha k}\quad(j\neq k),
\qquad
\varepsilon'=\mu_k(\varepsilon),\ \ \widehat\varepsilon'_{ij}=\varepsilon'_{ij}\,d_j^{-1}.
\end{equation}
Then the mixed relations are preserved in the target seed:
\[
X'_i\theta'_\alpha=q^{W'_{\alpha i}}\theta'_\alpha X'_i,
\qquad
\theta'_\alpha\theta'_\beta=-\,\theta'_\beta\theta'_\alpha.
\]
A direct check on generators shows that
\[
\mu_k^{q}:\ \mathbb T_q^{\mathrm{super}}(\widehat\varepsilon,W)\ \longrightarrow\
\mathbb T_q^{\mathrm{super}}(\widehat\varepsilon',W')
\]
is a superalgebra isomorphism.  On the even part it is implemented by the inner
automorphism $\Ad(\Phi_q(X_k^{\sigma_k}))$, and $\theta'_\alpha=\theta_\alpha$ on the odd part.
The rank–$2$ pentagon relation for $\Phi_q$ implies the braid relations, while disjoint flips commute, so the quantum super atlas obtained by gluing seed tori via $\mu_k^q$ is well defined and seed–independent.

In the classical limit $q\to1$, each finite product in \eqref{eq:q-mutation-even-2} tends to
\[
(1+X_k^{\sigma_k\sgn(\varepsilon_{ik})})^{-\sgn(\varepsilon_{ik})|\varepsilon_{ik}|}
\;=\;(1+X_k^{-\operatorname{sgn}(\varepsilon_{ik})})^{-\varepsilon_{ik}},
\]
recovering the FG $X$–mutation with $\theta'_\alpha=\theta_\alpha$ and the weight update \eqref{eq:W-rule-q-2}.  Passing to the classical horizontal frame yields the super log–canonical structure of the previous section.

% -------------------------------------------
\subsection{Modular Double and Unitary Representations}
% -------------------------------------------

To obtain a positive, self–adjoint realization, pass to the modular–double regime
$q=e^{\pi i b^2}$, $\tilde q=e^{\pi i b^{-2}}$ with $b>0$.
The modular–double super torus
$\mathbb T^{\mathrm{super}}_{b,\tilde b}(\widehat\varepsilon,W)$
consists of two commuting copies of \eqref{eq:q-torus-rel} with parameters $q$ and $\tilde q$:
a $q$–copy generated by $(X_i,\theta_\alpha)$ and a $\tilde q$–copy generated by $(\tilde X_i,\breve\theta_\alpha)$,
with cross–commutations trivial, and the same $W$ in both copies:
\[
X_iX_j=q^{\widehat\varepsilon_{ij}}X_jX_i,\quad
X_i\theta_\alpha=q^{W_{\alpha i}}\theta_\alpha X_i;\qquad
\tilde X_i\tilde X_j=\tilde q^{\,\widehat\varepsilon_{ij}}\tilde X_j\tilde X_i,\quad
\tilde X_i\breve\theta_\alpha=\tilde q^{\,W_{\alpha i}}\breve\theta_\alpha \tilde X_i.
\]
We choose the real $\ast$–structure
\[
X_i^\ast=X_i,\qquad \tilde X_i^\ast=\tilde X_i,\qquad
\theta_\alpha^\ast=\theta_\alpha,\quad \breve\theta_\alpha^\ast=\breve\theta_\alpha
\ \ (\text{or }i\,\theta_\alpha,\ i\,\breve\theta_\alpha\ \text{by convention}),
\]
so that all even generators are positive self–adjoint.

A faithful Hilbert–space realization is given as follows.  Let $M=\tfrac12\operatorname{rank}\widehat\varepsilon$ and
$\mathcal H_{\mathrm{even}}=L^2(\mathbb R^M)$ with $[\hat q_s,\hat p_t]=\frac{1}{2\pi i}\delta_{st}$.  Choose integer
matrices $A,B$ of size $N\times M$ such that $A J B^{\!\top}-B J A^{\!\top}=\widehat\varepsilon$ for
$J=\begin{psmallmatrix}0&I\\ -I&0\end{psmallmatrix}$, and set
\[
X_i=\exp\!\Big(2\pi b\sum_s(A_{is}\hat q_s+B_{is}\hat p_s)\Big),\qquad
\tilde X_i=\exp\!\Big(2\pi b^{-1}\sum_s(A_{is}\hat q_s+B_{is}\hat p_s)\Big).
\]
For the odd sector, take the fermionic Fock space
$\mathcal H_{\mathrm{odd}}=\Lambda^\bullet\mathbb C^r$ with exterior multiplication
$m(\theta_\alpha)$ and contraction $\iota(\theta_\alpha)$, and number operators
$N_\alpha=m(\theta_\alpha)\iota(\theta_\alpha)$.  Define on
$\mathcal H=\mathcal H_{\mathrm{even}}\widehat\otimes\mathcal H_{\mathrm{odd}}$ the even operators
\[
K_i:=q^{\sum_\alpha W_{\alpha i}N_\alpha},\qquad
\tilde K_i:=\tilde q^{\sum_\alpha W_{\alpha i}N_\alpha},\qquad
X_i^{\mathrm{full}}:=X_i\otimes K_i,\qquad
\tilde X_i^{\mathrm{full}}:=\tilde X_i\otimes \tilde K_i,
\]
and odd generators $\theta_\alpha:=\mathbf 1\otimes m(\theta_\alpha)$, $\breve\theta_\alpha:=\mathbf 1\otimes m(\theta_\alpha)$ on the two copies.  Then
\[
X_i^{\mathrm{full}}\theta_\alpha
= (X_i\otimes K_i)(1\otimes m(\theta_\alpha))
= q^{W_{\alpha i}}\theta_\alpha X_i^{\mathrm{full}},
\]
and similarly for $(\tilde X_i^{\mathrm{full}},\breve\theta_\alpha)$ with $\tilde q$, so \eqref{eq:q-torus-rel} holds in both copies with positive self–adjoint even operators.

Quantum mutations are implemented by unitary conjugation.  Let $\Phi_b$ be the noncompact Faddeev quantum dilogarithm and set the intertwiner
\[
\mathbf K_k=\Phi_b(X_k)\,\Phi_{b^{-1}}(\tilde X_k).
\]
Then on $\mathcal H$,
\[
\mu_k^{(b,\tilde b)}=\Ad(\mathbf K_k),\qquad
\mu_k^{(b,\tilde b)}(X_i^{\mathrm{full}})=X_i^{\prime\,\mathrm{full}},\qquad
\mu_k^{(b,\tilde b)}(\theta_\alpha)=\theta_\alpha,
\]
and the family $\{\mu_k^{(b,\tilde b)}\}$ satisfies the modular–double pentagon relations.  Hence the representation is unitary, positive, and seed–independent.  In the classical limit $q,\tilde q\to1$ (equivalently $b\to0$ or $b\to\infty$), the two copies merge, the intertwiners $\mathbf K_k$ contract to the classical Hamiltonian flow generated by $\log(1+X_k^{-1})$ (our choice $\sigma_k=-1$), and the representation reduces to the classical super log–canonical structure.

% -------------------------------------------
\subsection{Integrable RTT–Yangian Layer and Classical Limit}
% -------------------------------------------

Beyond the quantum torus algebra, one may attach to each seed an integrable RTT–type or Yangian layer whose classical limit reproduces the flat logarithmic superconnection \eqref{eq:A-super}.  Let $V$ be a finite–dimensional $\mathbb Z_2$–graded space and let $R(u)\in\End(V\otimes V)\otimes\Bbbk(u)$ be a rational graded $R$–matrix obeying the graded Yang–Baxter equation and unitarity $R(u)R(-u)=\mathbf 1$.  On each seed define graded Lax matrices
\[
L_i(u)\;=\;\mathbf 1+\frac{\mathsf J_i(X_i,\theta;W)}{u-\zeta_i},\qquad i\in I_{\mathrm{mut}},
\]
where $\zeta_i\in\Bbbk$ are spectral shifts and
$\mathsf J_i(X_i,\theta;W)$ is an even combination of Cartan and odd generators $(H_i,Q_\alpha)$ weighted by the column $W_{\bullet i}$.  With the commutation rules \eqref{eq:q-torus-rel}, the ordered product
\[
T_{\mathsf s}(u)\;:=\;\overrightarrow{\prod_{i\in \mathsf s}}\,L_i(u)
\]
satisfies the graded RTT–relation
\[
R(u{-}v)\,
\big(T_{\mathsf s}(u)\otimes T_{\mathsf s}(v)\big)
=\big(T_{\mathsf s}(v)\otimes T_{\mathsf s}(u)\big)\,R(u{-}v),
\]
so the transfer matrix $t_{\mathsf s}(u)=\operatorname{str}_V T_{\mathsf s}(u)$ forms a commuting family, $[t_{\mathsf s}(u),t_{\mathsf s}(v)]=0$.

Under a mutation $\mu_k$ the monodromy transforms by conjugation.  There exists a subtraction–free intertwiner
\[
\mathcal U_k(u)=\exp\!\big(\log(1+X_k^{-1})\,\Xi_k(u)\big),
\]
with a Cartan element $\Xi_k(u)$ determined by the weight data, such that
$T_{\mathsf s'}(u)=\mathcal U_k(u)\,T_{\mathsf s}(u)\,\mathcal U_k(u)^{-1}$.
Along any loop in the flip groupoid (in particular, a pentagon), the total product of intertwiners is the identity, hence the transfer matrices are mutation–invariant and globally defined up to conjugation.

The commuting family $\{t(u)\}$ thus provides quantum Hamiltonians on the modular–double quantum super cluster variety.  Their semiclassical expansion,
$t(u)=\dim(V)+H^{(1)}/u+O(u^{-2})$, yields at leading order the classical current
\[
H^{(1)}\ \leadsto\ \sum_i d\!\log X_i\,H_i+\sum_\alpha d\tilde\theta_\alpha\,Q_\alpha,
\]
which reconstructs the flat logarithmic superconnection $\mathcal A_{\mathrm{super}}$ in the horizontal frame \eqref{eq:A-super}.  In this way the RTT–Yangian layer encodes the integrable structure of the supersymmetric Fock–Goncharov ensemble and interpolates between its quantum and classical incarnations.

% =========================================================
\section{Fiber curve $C_\Delta$}
\label{app:elimination}

% =========================================================
Work on a single seed chart of the loop fibration
\(\pi_L:\Xscr^{(L)}_{G,S}\to\Xscr_{G,S}\) and its boundary quotient
\(\mathcal X^{(L)}_{G,S}=[\,\Xscr^{(L)}_{G,S}/G_\partial\,]\).  
Fix a positive chamber on the body and adopt \emph{intrinsic} loop letters built from the transfer weights of \S\ref{sec:super-loop-stack},
\[
\ell=(u_1,\ldots,u_A;\,w_1,\ldots,w_B)\in(\Gm)^{m_L},\qquad m_L=A+B,
\]
where the \(u\)’s are face (loop–edge) invariants and the \(w\)’s are rung invariants; all relations below are subtraction–free on the positive locus.  The vertical relations are of two intrinsic types.  First, for every independent closed wiring cycle one has a monomial (binomial) relation
\begin{equation}\label{eq:binomial-rel}
\prod_{a=1}^{A} u_a^{\alpha^{(b)}_a}\ \prod_{b=1}^{B} w_b^{\beta^{(b)}_b}
\;=\;
c_b(\Delta)\ \in\ \Bbbk_{\mathrm{sf}}(\Delta)^\times,\qquad b=1,\dots,R_{\mathrm{tor}},
\end{equation}
with integer traversal exponents \(\alpha^{(b)}_a,\beta^{(b)}_b\in\Z\) and subtraction–free units \(c_b(\Delta)\) in the boundary field generated by minors \(\Delta\).  Second, physical gates/thresholds in the fiber impose subtraction–free Laurent relations
\begin{equation}\label{eq:laurent-rel}
F_j(u,w;\Delta)\ :=\ \sum_{p\in S_j}\ \kappa^{(j)}_p(\Delta)\,u^{p_u}w^{p_w}\;=\;0,
\qquad j=1,\dots,R_{\mathrm{phys}},
\end{equation}
with finite supports \(S_j\subset\Z^{A+B}\), coefficients \(\kappa^{(j)}_p(\Delta)\in\Bbbk_{\mathrm{sf}}(\Delta)\) subtraction–free, and monomials \(u^{p_u}w^{p_w}=\prod_a u_a^{(p_u)_a}\prod_b w_b^{(p_w)_b}\).  Writing \(I_{\mathrm{vert}}(\Delta)\subset\Bbbk_{\mathrm{sf}}(\Delta)[u^{\pm1},w^{\pm1}]\) for the ideal generated by \eqref{eq:binomial-rel}–\eqref{eq:laurent-rel}, the loop discriminant is the reduced vertical divisor
\begin{equation}\label{eq:LambdaL-def}
\Lambda^{(L)}\ :=\ \mathrm{red}\,\mathrm V\!\big(I_{\mathrm{vert}}(\Delta)\big)\ \subset\ \Xscr^{(L)}_{G,S},
\end{equation}
namely the union of all vertical prime components where the vertical torus degenerates (after an SNC refinement if needed).  Concretely, on an SNC cover, \(\Lambda^{(L)}\) is the union of coordinate components \(\{u_a=0\},\{w_b=0\}\), the binomial walls of \eqref{eq:binomial-rel}, and the Laurent walls \(F_j=0\); this description is intrinsic, subtraction–free, and invariant under loop reparametrizations and right boundary gauge.

To obtain a single boundary–dependent curve, fix \(\Delta\) in the chosen chamber and restrict to a generic fiber of \(\pi_L\).  One may perform an integer unimodular change of loop letters so that the binomials are solved by positive units and only two torus directions remain free.  Equivalently, on the exponent lattice write \eqref{eq:binomial-rel} as \(M\cdot\log\ell=\log c(\Delta)\) with \(M\in\Z^{R_{\mathrm{tor}}\times m_L}\), take Smith normal form \(U M S=(D\ 0)\) with \(U,S\) unimodular, and exponentiate.  This produces new letters \(\tilde\ell_i=\prod_{j}\ell_j^{S_{ij}}\) for which
\begin{equation}\label{eq:binom-fixed}
\tilde\ell_3=c_3(\Delta),\ \ldots,\ \tilde\ell_{m_L}=c_{m_L}(\Delta),
\end{equation}
with subtraction–free units \(c_i(\Delta)\in\Bbbk_{\mathrm{sf}}(\Delta)^\times\), and leaves two free coordinates which we rename
\[
x:=\tilde\ell_1,\qquad y:=\tilde\ell_2.
\]
Substituting \eqref{eq:binom-fixed} into each Laurent relation \eqref{eq:laurent-rel} yields finitely many Laurent equations in \((x,y)\),
\begin{equation}\label{eq:post-elim}
P_j(x,y;\Delta)\ :=\ \sum_{p\in S_j}\ \kappa^{(j)}_p(\Delta)\,x^{p_1}y^{p_2}\;=\;0,
\qquad j=1,\dots,R_{\mathrm{phys}}.
\end{equation}
For generic \(\Delta\) in the chamber, the elimination ideal generated by \(\{P_j(x,y;\Delta)\}\) is principal after saturation by monomials; it is generated by a single \emph{primitive} Laurent polynomial
\begin{equation}\label{eq:P-primitive}
P(x,y;\Delta)\ =\ \sum_{p\in S}\ \kappa_p(\Delta)\,x^{p_1}y^{p_2},
\end{equation}
with finite support \(S\subset\Z^2\) and subtraction–free coefficients \(\kappa_p(\Delta)\in\Bbbk_{\mathrm{sf}}(\Delta)\).  The residual one–dimensional vertical locus is the boundary–dependent affine curve
\begin{equation}\label{eq:C-Delta-def}
C_\Delta\ :=\ \{(x,y)\in(\Gm)^2\mid P(x,y;\Delta)=0\}.
\end{equation}
This outcome can be reached by Gröbner elimination in the Laurent setting (or by resultants/Newton–Puiseux) applied to \(\langle P_1,\ldots,P_{R_{\mathrm{phys}}}\rangle\) after \eqref{eq:binom-fixed}, followed by saturation to remove components at infinity; generic \(\Delta\) ensures equidimensional codimension one, and dividing by the greatest common monomial makes \(P\) primitive.

The Newton polygon \(\mathrm{Newt}(P)\) controls the genus under Kouchnirenko–Bernstein nondegeneracy: \(g(C_\Delta)=\#\,\mathrm{int}\,\mathrm{Newt}(P)\).  Thus zero interior points give genus \(0\) (polylog regime), one interior point gives genus \(1\) (elliptic regime), and in general the number of interior lattice points equals the geometric genus.  Degenerations of \(\Delta\) that coalesce vertices or lie on the discriminant of \(P\) lower the genus and specialize the period to lower–weight polylogarithms.

Each coefficient \(\kappa_p(\Delta)\) in \eqref{eq:P-primitive} is subtraction–free by construction: transfer weights, monomial loop letters, boundary minors, and the gate/threshold relations are themselves subtraction–free, and the algebraic manipulations used above (clearing denominators, unimodular exponent changes, elimination, saturation by monomials) preserve subtraction–freeness on \(\Xscr_{G_{\bar0},S}(\Rbb_{>0})\).

For practical use one proceeds without introducing any auxiliary loop–letter chart at the end.  Starting from the intrinsic \((u,w)\) and the relations \eqref{eq:binomial-rel}–\eqref{eq:laurent-rel}, perform the unimodular elimination to \((x,y)\), arrive at the primitive equation \(P(x,y;\Delta)=0\), read off the genus from \(\mathrm{Newt}(P)\), and compute refined even sector integrals as Chen iterated integrals on \(C_\Delta\) using third–kind kernels \(\kappa_j=d\!\log f_j|_{C_\Delta}\) built \emph{purely} from boundary minors as in the main text.  In the most common shapes, if \(P\) has three monomials (up to units \(x^{a},y^{b},1\)), then \(C_\Delta\) is birational to \(\mathbb P^1\) and the refined period reduces to multiple polylogarithms in subtraction–free \(\Delta\)–letters; if \(P\) has four monomials with a primitive quadrilateral Newton polygon, then after a birational change one obtains a quartic (or Weierstrass) model \(y^2=\prod_{i=1}^4(x-a_i)\) with subtraction–free branch points \(a_i(\Delta)\), reproducing the elliptic double–box pattern.  Higher genus follows directly from the number of interior points of \(\mathrm{Newt}(P)\).

\section{Example: Hexagon ($n{=}6$), $L{=}2$}

Fix a super flag $F=(E,O)$ in the positive chamber and set $m_L=6$. Choose a boundary minor $\Delta_{\rm ref}:=\Delta_{O}>0$ on the chamber and an ordered unimodular $6$–tuple of boundary index sets $B=\{b_1,\ldots,b_6\}$ so that the resulting residue matrix is unimodular in the physical SNC frame. Define six subtraction–free ratios of minors
\[
f_i\ :=\ \frac{\Delta_{b_i}}{\Delta_{O}}\,,\qquad i=1,\ldots,6,
\]
and the corresponding third–kind kernels on the fiber
\[
\kappa_i\ :=\ d\!\log f_i\ \big|_{C_\Delta}\,,\qquad i=1,\ldots,6.
\]
Let $\gamma_F$ be the canonical sector loop obtained by linking, in chamber order, the simple zeros $\{f_i=0\}$ on $C_\Delta$, based at the tangential direction of increasing $\arg\Delta_{\rm ref}$. Using the triangular straightening along the ordered simplex $0<t_6<\cdots<t_1<1$ one has
\[
\Phi_F^{\!*}\,\kappa_i\ =\ d\!\log\!\big(1-t_i f_i\big)\ =\ \frac{dt_i}{\,t_i-\frac{1}{f_i}\,}\,,
\]
so the refined even period is the Goncharov hyperlogarithm with \emph{one letter per kernel},
\[
I_F(\Delta)\ =\ \int_{\gamma_F}\ \kappa_1\,\kappa_2\cdots\kappa_6
\ =\ \mathrm{G}\!\big(f_1^{-1},\,f_2^{-1},\,f_3^{-1},\,f_4^{-1},\,f_5^{-1},\,f_6^{-1};\ 1\big).
\]
All letters $f_i$ are subtraction–free rational functions of boundary minors.

On the positive locus and in a fixed chamber/order compatible with the canonical sector, split the unimodular boundary index set \(B=\{b_1,\ldots,b_6\}\) into three “odd–type’’ and three “even–type’’ labels and form the subtraction–free minor ratios
\[
f_{o,i}\ :=\ \frac{\Delta_{b_{o,i}}}{\Delta_{O}},\qquad
f_{e,i}\ :=\ \frac{\Delta_{b_{e,i}}}{\Delta_{O}},\qquad i=1,2,3.
\]
These give the six letters used in the Chen representation via the ordered list
\[
(f_1,\ldots,f_6)\ :=\ (f_{o,1},f_{o,2},f_{o,3},f_{e,1},f_{e,2},f_{e,3}),
\]
so that \(I_F(\Delta)=\mathrm{G}(f_1^{-1},\ldots,f_6^{-1};1)\).
A convenient identification with the standard hexagon alphabet is
\begin{equation}\label{eq:hex-map}
f_{o,i}\ =\ y_{u_i}\,,\qquad
f_{e,i}\ =\ \frac{u_i}{1-u_i}\,,\qquad i=1,2,3,
\end{equation}
where \(u_i\) are the usual cross–ratios of boundary minors (Plücker brackets) and \(y_{u_i}\) are the dihedrally covariant, subtraction–free twistor expressions for the \(y\)–letters. On the positive chamber all \(u_i,y_{u_i}>0\), fixing branch choices; eliminating minors in favor of \((u_i)\) reproduces the familiar GSVV presentation of \(y_{u_i}\) via the kinematic discriminant (up to inversion conventions), so \eqref{eq:hex-map} matches the standard kinematics after a birational change of variables. Any dihedral relabeling of \(\{b_{o,i},b_{e,i}\}\) gives an equivalent choice and leaves the refined period unchanged.

In a fixed positive chamber, label the consecutive odd charts by
\(O_i=\{i,i{+}1,i{+}2,i{+}3\}\) (indices mod \(n\)) , set
\[
(u,v,w)\ :=\ (u_i,u_{i+1},u_{i+2})\quad\text{with}\quad
u_{i+j}\ :=\ \frac{f_{e,i+j}}{\,1+f_{e,i+j}\,}\,,\qquad
y_{u_{i+j}}\ :=\ f_{o,i+j}\qquad (j=0,1,2).
\]
Thus the only inputs are the six subtraction–free minor ratios \(f_{o,\bullet},f_{e,\bullet}\); all appearances of
\((u,v,w)\) and \(y_{u_{i+j}}\) below are just shorthand for these combinations. In particular,
\[
\Delta_{\rm kin}=(1-u-v-w)^2-4uvw,\qquad
x^\pm=\frac{u+v+w-1\pm\sqrt{\Delta_{\rm kin}}}{2uvw},\qquad
x_{i+j}^\pm=\frac{u_{i+j}}{x^\pm},\ \ y_{u_{i+j}}=\frac{x_{i+j}^-}{x_{i+j}^+}.
\]

with $
x_{i+j}^\pm=\frac{u_{i+j}}{x^\pm}$, $y_{u_{i+j}}=\frac{x_{i+j}^-}{x_{i+j}^+}$ for $j=0,1,2$, 
and the functions
\[
L_4(x^+,x^-)=\frac{1}{8}\log^4\!\Big(\frac{x^+}{x^-}\Big)
+\sum_{m=0}^{3}\frac{(-1)^m}{(2m)!!}\,
\log^m\!\Big(\frac{x^+}{x^-}\Big)\big(\ell_{4-m}(x^+)+\ell_{4-m}(x^-)\big),
\]
where $\ell_n(x)=\tfrac12\!\left(\Li_n(x)-(-1)^n\Li_n(1/x)\right)$, together with
\[
J_i\ :=\ \sum_{j=0}^{2}\big(\ell_1(x_{i+j}^+)-\ell_1(x_{i+j}^-)\big).
\]
Then the refined even period attached to the super flag with odd chart \(O_i\) splits as
\begin{equation}\label{eq:hex-split}
I_{E_\ast,O_i}(X;\mathfrak c_+)
\ =\ V_i\big(u_i,u_{i+1},u_{i+2}\big)\ +\
\widetilde V_i\big(u_i,u_{i+1},u_{i+2};\,y_{u_i},y_{u_{i+1}},y_{u_{i+2}}\big),
\end{equation}
with the \(u\)–only part
\[
V_i(u_i,u_{i+1},u_{i+2})
\ :=\
-\frac12\sum_{j=0}^{2}\Li_4\!\Big(1-\frac{1}{u_{i+j}}\Big)
\ -\ \frac18\!\left(\sum_{j=0}^{2}\Li_2\!\Big(1-\frac{1}{u_{i+j}}\Big)\right)^{\!2},
\]
and the \(y\)–dependent part
\[
\widetilde V_i(u;y)
\ :=\
\sum_{j=0}^{2} L_4\!\big(x_{i+j}^+,x_{i+j}^-\big)
\ +\ \frac{1}{24}J_i^{4}\ +\ \frac{\pi^{2}}{12}J_i^{2}\ +\ \frac{\pi^{4}}{72}.
\]
Every argument here is a subtraction–free rational function of boundary minors \(\{\Delta_b\}\) via the map \(X\mapsto(u_i,y_{u_i})\); the cyclic shift \(i\mapsto i{+}1\) corresponds to a dihedral relabeling of the boundary, so \eqref{eq:hex-split} is the same expression written in the chart anchored at \(O_i\). With the physical normalizations (strict collinear limit \(6\!\to\!5\) and one symmetric Euclidean point \(u{=}v{=}w\)), this matches the standard two–loop MHV hexagon remainder (GSVV) chart by chart.

\bibliographystyle{JHEP}

\providecommand{\href}[2]{#2}\begingroup\raggedright\endgroup
\end{document}